\documentclass[a4paper,fleqn,usenatbib]{mnras}

\usepackage{amsfonts}

\usepackage[T1]{fontenc}
\usepackage{ae,aecompl}

\usepackage{graphicx}	
\usepackage{amsmath}	
\usepackage{amssymb}	
\usepackage{caption}
\usepackage{float}
\usepackage{lscape}

\newcommand{\teff}{\mbox{$T_{\rm eff}$}}
\newcommand{\logg}{\mbox{$\log g$}}
\newcommand{\halpha}{\mbox{$\mathrm{H} \alpha$}}

\newcommand{\mictrb}{\mbox{$v_{\rm mic}$}}
\newcommand{\loggf}{\mbox{$\log gf$}}
\newcommand{\kms}{\mbox{km\,s$^{-1}$}}

\title[HoSTS II]{The Homogenous Study of Transiting Systems (HoSTS). II. The influence of the line list on stellar parameters}

\author[A.P. Doyle et al.]{
Amanda P. Doyle,$^{1}$\thanks{E-mail: Amanda.Doyle@warwick.ac.uk}
Barry Smalley,$^{2}$
Francesca Faedi,$^{1}$
Don Pollacco,$^{1}$ \newauthor
Yilen G{\'o}mez Maqueo Chew$^{3}$
\\
$^{1}$Department of Physics, University of Warwick, Gibbet Hill Road, Coventry
CV4 7AL, UK\\
$^{2}$Astrophysics Group, Keele University, Staffordshire ST5 5BG, UK\\
$^{3}$Instituto de Astronom\'ia, Universidad Nacional Aut\'onoma de M\'exico, Ciudad Universitaria, 04510, Ciudad de M\'exico, M\'exico
}

\date{Accepted XXX. Received YYY; in original form ZZZ}

\pubyear{2016}

\begin{document}
\label{firstpage}
\pagerange{\pageref{firstpage}--\pageref{lastpage}}
\maketitle

\begin{abstract}
The use of high resolution, high signal-to-noise stellar spectra is essential in order to determine the most accurate and precise stellar atmospheric parameters via spectroscopy. This is particularly important for determining the fundamental parameters of exoplanets, which directly depend on the stellar properties. However, different techniques can be implemented when analysing these spectra which will influence the results. These include performing an abundance analysis relative to the solar values in order to negate uncertainties in atomic data, and fixing the surface gravity (\logg ) to an external value such as those from asteroseismology. The choice of lines used will also influence the results. In this paper, we investigate differential analysis and fixing \logg\ for a set of FGK stars that already have accurate fundamental parameters known from external methods. We find that a differential line list gives slightly more accurate parameters compared to a laboratory line list, however the laboratory line list still gives robust parameters. We also find that fixing the \logg\ does not improve the spectroscopic parameters.  We investigate the effects of line selection on the stellar parameters and find that the choice of lines used can have a significant effect on the parameters. In particular, removal of certain low excitation potential lines can change the \teff\ by up to 50 K.  For future HoSTS papers we will use the differential line list with a solar microturbulence value of 1 \kms, and we will not fix the \logg\ to an external value.

\end{abstract}

\begin{keywords}
stars: abundances, stars: fundamental parameters
\end{keywords}



\section{Introduction}
Our understanding of exoplanets is strongly tied to our understanding of their host stars. The mass and radius of a transiting exoplanet cannot be determined without first knowing the mass and radius of the star \citep{Winn11}. Direct determinations of stellar mass and radius are only possible for those in a binary system \citep{Andersen91}. For other stars, the mass and radius needs to be determined indirectly. This can be done via asteroseismology once the effective temperature (\teff) is known \citep{Chaplin11}. The planet transit can yield the stellar density, which can be used to determine the stellar mass and radius once the \teff\ and metallicity are known \citep{Sozzetti07}. The distance to a star, which can be measured via parallax with \textit{Gaia}, can also be used to determine the stellar radius, once \teff\ and bolometric flux are known \citep{Stassun16}. 

For stars without asteroseismic measurements, spectroscopy is used to determine the \teff, surface gravity (\logg), and metallicity ([Fe/H]). These parameters are then input into a calibration (e.g. \citet{Torres10}) or used with a grid of stellar models (e.g. \citet{Girardi00}) to find the stellar mass and radius.

For the spectroscopic analysis, some authors rely on differential analysis as opposed to the atomic data from the VALD database (e.g.  \citet{Melendez09}, \citet{Bruntt10}, \citet{Brugamyer11}, \citet{Sousa14}). The disadvantage is that differential analysis can only be performed accurately for stars with similar parameters to the Sun. However, while it is clear that stars with properties that deviate vastly from solar should not be used in differential analysis, it is not clear what the cut-off in parameters should be for planet host stars, which are typically FGK dwarfs and subgiants, although planets have also been found around giant stars (e.g. \citet{Wittenmyer17}).  In this paper we analyse a set of 23 FGK stars and compare a line list using the laboratory data from VALD with a differential line list in order to investigate if the differential list is really the best one to use and if errors are more likely for stars with parameters furthest from solar.   

Due to the uncertainties inerrant in the spectroscopic \logg, fixing the \logg\ to an external value such as those obtained from the planet transit or asteroseismology can improve the spectroscopic parameters for some methods which directly compare the observed spectrum with a synthetic spectrum. For example, \citet{Torres12} found that fixing \logg\ improved the parameters when using the software {\sc SME} and {\sc SPC}, but when they used an equivalent width (EW) based method with the software MOOG, they found the latter to be more accurate. \citet{Mortier14} also found the EW-based method does not require \logg\ to be fixed to an external value.

The Homogeneous Study of Transiting Systems (HoSTS) project aims to characterise planets and their host stars consistently, and to use a homogeneous, high quality set of stellar spectra. Four spectral analysis methods were compared in the pilot study of WASP-13 (\citet{Gomez13}; hereafter Paper I) using a high resolution (R = 72~000) HIRES spectrum. Each of the methods performed three different analyses: an unconstrained analysis to obtain \teff, \logg\ and [Fe/H], an analysis with the \teff\ fixed to that determined from the \halpha\ line from a long-slit IDS spectrum, and an analysis with \logg\ fixed from the transit value. Paper I found that the results from the unconstrained analysis agreed well between all four methods and are consistent with the transit \logg\ and the \halpha\ \teff, leading to the conclusion that the four different methods of spectral analysis have no systematic differences between them.

In this paper we investigate differential analysis and fixing the \logg\ to an external value, in order to determine if there is any preferential method. We also determine the effect that line selection will have on the stellar parameters. We use the Fe-line method, where the EWs are measured for a number of Fe lines in order to determine the stellar parameters. We also obtain the \logg\ from the pressure broadened Mg~{\sc i} b and Na~{\sc i} D lines.  

The stars that we chose to analyse are a set of 23 bright, standard stars that were previously analysed by \citet{Bruntt10} (hereafter B10). This selection of standard stars was chosen for this work as the parameters span the range of stars that can host exoplanets, i.e. FGK stars from dwarfs to giants. B10 determined non-spectroscopic parameters for these stars, which are a useful test of our spectroscopic parameters; the bolometric \teff, photometric \teff, the \logg\ determined from a binary mass and interferometric angular diameter, and asteroseismic \logg. The bolometric \teff\ from \citet{Heiter15} (hereafter H15) is also used.

All of the spectra that we used were taken from the ESO HARPS archive and the spectra all have S/N $\sim$100. While higher S/N could be achieved by coadding several spectra, a S/N of 100 is more representative of a typical planet host star. 

In Section~\ref{Methods}, we discuss the methods used for our analysis and the selection of the line lists. Section~\ref{Results} details our results where the results between the VALD and differential line lists are compared. Section~\ref{Discussion} discusses the results and we conclude in Section~\ref{Conclusion}.

\section{Methods}
\label{Methods}

\subsection{Measuring spectral parameters}
The spectroscopic parameters can be determined using a set of Fe lines and we use the same method as in \citet{Doyle13}. Once the equivalent widths (EWs) of the lines have been measured, the abundance is calculated for each line. The low excitation potential (EP) Fe~{\sc i} lines are sensitive to temperature, where as the temperature sensitivity is negligible for high EP lines. Thus requiring that there is no trend between EP and abundance will yield the \teff\ of the star and this is known as the excitation balance \teff. The error for the excitation balance \teff\ is from the 1-$\sigma$ variation in the slope of abundance against excitation potential. The same principle can be applied to Fe~{\sc ii} lines, but in this case it is the high EP lines that are sensitive to \teff\ changes. However, there are usually an insufficient number of Fe~{\sc ii} lines present in solar-like stars to determine the \teff. 

The \logg\ of the star can be determined via the ionisation balance, which occurs when the Fe~{\sc i} and Fe~{\sc ii} abundances agree. This is because the Fe~{\sc ii} abundance will increase with increasing \logg, where as the Fe~{\sc i} abundance is insensitive to \logg\ variations \citep{Takeda02a}. The error for the ionisation balance \logg\ is determined by varying the \teff\ by 1$\sigma$. It should also be noted that the number of Fe~{\sc ii} lines used is an important factor in determining the ionisation balance \logg. An insufficient number of Fe~{\sc ii} lines will lower the \logg, and likely explains the low \logg\ found in Paper I for WASP-13 using the \citet{Schuler11} line list, as this list has only 5 Fe~{\sc ii} lines.

The microturbulence (\mictrb) is a line broadening parameter required in 1D analyses that also affects abundance and thus the derived \teff\ and \logg. Microturbulence was introduced by \citet{Struve-Elvey34} so that the abundance calculated from strong lines would be the same as for weak lines. Therefore the \mictrb\ is determined by requiring that there is no slope between the abundance and EW.

The determination of parameters via the Fe lines is an iterative process. The excitation balance \teff\ is first determined using an initial \teff\ guess from the bolometric, photometric, or the B10 spectroscopic value. An initial guess of 1 \kms\ is used for the \mictrb. A \teff\ value is determined when the slope of the excitation balance plot is zero. The \logg\ is then adjusted so that the Fe~{\sc i} and Fe~{\sc ii} abundances agree, and the \mictrb\ is adjusted until there is no trend between abundance and EW. The Fe~{\sc i} lines used to determine the excitation balance \teff\ have very little dependence on the \logg\, so the \logg\ will not change the excitation balance. However, \teff\ is sensitive to \mictrb, which will change the slope so that excitation balance no longer occurs. A second iteration of \teff\ is therefore performed using the new \mictrb\ value. The new \teff\ will affect the \logg, therefore the ionisation balance needs to be redetermined. Several iterations are performed until the slopes of both plots have been minimised and the ionisation balance is correct.

\subsection{Software}

{\sc uclsyn} (University College London SYNthesis) is the software we used for spectral analysis (\citealt {Smith92}; \citealt {Smalley01}). {\sc atlas} 9 models without convective overshooting were used \citep{Castelli97} and local thermodynamic equilibrium (LTE) is assumed. Spectral lines were measured manually by using equivalent widths. A global continuum fit was performed using iSpec \citep{Blanco-Cuaresma14}, however the local continuum still needed to be adjusted on a line-by-line basis within {\sc uclsyn}. The atomic data used to generate the synthetic spectra was obtained from \citet {Kurucz-Bell95}, although lines can also be input manually using atomic data from other sources. The broadening parameters that are input are microturbulence, macroturbulence, rotational velocity, and instrumental broadening. The radiative damping constant, Van der Waals damping constant, and the Stark broadening factor are input via the line list. We used the solar abundances from \citet{Asplund09}.

\subsection{Line list}
\subsubsection{Line selection}
We selected spectral lines from the Kitt Peak Solar Atlas \citep{Kurucz84} by looking for unblended lines. It is important to have as many low EP lines as possible for the determination of \teff\ via the excitation balance, however many of these lines are in ``resolved blends'' \footnote{A resolved blend is where two close-by spectral lines have blended wings, but the individual lines can still be measured.}, which meant that they were initially ignored when selecting lines from the solar spectrum. All of the low EP lines listed in the NIST database \citep{Fuhr-Wiese06} were then checked in the solar spectrum. Any that were still measurable despite being in a resolved blend were added to the line list. Any lines with EW greater than 0.12 {\AA} were not included as these will skew the \mictrb. This is because the stronger lines are affected more by \mictrb\ and also because the stronger lines are more difficult to measure due to the extent of the wings.

A line selected in the Sun may not necessarily be measurable in other stars. Stars with higher metallicity will have more blended lines, as will cooler stars. However, as we checked each line manually in each star, it is not necessary to have different line lists for different \teff\ and metallicity ranges for our method. The majority of the lines are measurable in all of our spectra. 

\subsubsection{Low metallicity}
Two of the stars in this sample have low metallicity; 171 Pup has [Fe/H] = $-$0.76, and $\nu$ Ind has [Fe/H] = $-$1.46. It is still possible to get a solution for these stars, although most of the lines are very weak so are more prone to errors. Ideally a separate line list should be used for low metallicity stars so as not to use only very weak lines. Lines which are easily measurable and unblended in low metallicity stars are usually too strong or blended in solar-metallicity stars. However, as most planet host stars do not have such low metallicities, we deemed it beyond the scope of this work to also create a line list for the low metallicity stars.

\subsubsection{Atomic data}
The oscillator strength (\loggf) will affect the derived abundance for a line, however these values are not always known to great accuracy, which can create large errors in abundance. In order to deal with this problem, some authors instead use differential \loggf\ values, i.e. those that are normalised to the Sun. Here, we create three line lists; one using the laboratory values thus giving the absolute abundances, and two differential line lists that use two different solar \mictrb\ values.

For the laboratory line list, the atomic data were taken from the VALD III  database \citep{Ryabchikova15} and only the laboratory values for \loggf\ were used. VALD II data were used for the \loggf\ values for Fe~{\sc ii}. This is because the Fe~{\sc ii} \loggf\ values from VALD III result in Fe~{\sc ii} abundances that are too low in the Sun. This changes the ionisation balance \logg, which in turn changes the \mictrb. As \teff\ is dependent on \mictrb, the resulting \teff\ for the Sun becomes 5701 $\pm$ 33 K. The \logg\ and \mictrb\ are 4.48 $\pm$ 0.08 dex and 0.72 $\pm$ 0.03 \kms\ respectively. This \teff\ is too low compared to the known solar value of 5777 K \citep{Gray08}, however when using the Fe~{\sc ii} \loggf\ values from VALD II \citep{Kupka99} the solar parameters are now 5750 $\pm$ 31 K, 4.42 $\pm$ 0.06 dex, and 0.83 $\pm$ 0.02 \kms\ for \teff, \logg, and \mictrb\ respectively. 

It is known that the VALD values can be poorly determined, resulting in a large ($\sim$0.6 dex) dispersion even in the Sun \citep{Sousa14}. In order to have a line list that uses the VALD atomic data and still be as accurate as possible, we imposed several criteria for line selection. If a selected line did not have laboratory data available, then it wasn't used. All lines were also required to have an abundance that didn't deviate more than 0.25 dex from the mean abundance in the Sun. When these lines were used across all of the sample stars, we found that there were still a number of lines that resulted in abundances that were consistently too low or too high in all stars compared to the mean abundance of each star. If there was an alternate laboratory \loggf\ available in VALD that gave a better abundance, then this was used. If there were no alternate values available, then the lines were deleted from the list. Our $\log A$(Fe) for the Sun is 7.56 $\pm$ 0.07\footnote{Abundances can be given in the format of $\log(A)$ + 12, where $\log(A)$ is the number ratio of the element with respect to hydrogen, log~($N_{el}$/$N_{H}$). The format [X/H] refers to the abundance of an element relative to the Sun, i.e. [Fe/H] = 0 for the Sun.} and the metallicity error is from the dispersion in the abundance values. 

For the two differential line lists, the \loggf\ was adjusted so that all of the lines give the \citet{Asplund09} abundance of $\log A$(Fe) = 7.5 in the Sun. In order to calculate the abundance of each line, the solar parameters (\teff\ = 5777 K and \logg\ = 4.44 dex) need to be input, along with the solar \mictrb. Two different lists were produced using two different assumptions of solar \mictrb; 1.0 and 0.85 \kms, as there is no standard value for the solar \mictrb. Throughout the paper, the two lists will be referred to as differential(1) and differential(0.85). The line list is given in the appendix and uses the differential \loggf\ values with a solar \mictrb\ of 1 \kms.

\section{Results}
\label{Results}

The parameters from B10 and H15 are given in Table~\ref{Bruntt}. The \teff, \logg, metallicity, and \mictrb\ results from this work are displayed in Tables~\ref{tefftable}, \ref{loggtable}, \ref{FeHtable}, and \ref{micttable} respectively. 

\subsection{Comparison between laboratory and differential line lists}

\begin{table*}
\centering
\begin{minipage}{140mm}
	\caption{The bolometric and photometric \teff, the binary and asteroseismic \logg, and the [Fe/H] from \citet{Bruntt10} are listed. The errors on the photometric \teff\ are 90 K and the errors on the [Fe/H] are 0.07 dex. The bolometric \teff\ values from \citet{Heiter15} are also given. }
	\label{Bruntt}
	\begin{tabular}{lllllllr} 
		\hline
Star name 		& HD 			& 	$\teff_{\rm bol}$ 	 & $\teff_{\rm bol}$	 & $\teff_{\rm phot}$  &  $\logg_{\rm bin}$ & $\logg_{\rm ast}$ & [Fe/H] \\
 & & B10 & H15 & & & &  \\
\hline
171 Pup			& 63077 		& 		 &  		 & 5790  &  		    & 4.244 $\pm$ 0.023 & $-$0.86 \\
70 Oph A		& 165341		& 		 &		 &       & 4.468 $\pm$ 0.030  & 4.555 $\pm$ 0.023 & 0.12  \\
$\alpha$ Cen A 		& 128620 	 	& 5746 $\pm$ 50  & 5792 $\pm$ 16 & 5635  & 4.307 $\pm$ 0.005  & 4.318 $\pm$ 0.017 & 0.22  \\
$\alpha$ Cen B 		& 128621 		& 5140 $\pm$ 56  & 5231 $\pm$ 20 &       & 4.538 $\pm$ 0.008  & 4.530 $\pm$ 0.018 & 0.30 \\
$\alpha$ For 		& 20010 		& 		 & 		 & 6105  & 		    & 4.003 $\pm$ 0.033 & $-$0.28 \\
$\alpha$ Men 		& 43834 		& 		 & 		 &       & 		    & 	    & 0.15             \\
$\beta$ Aql	 	& 188512 		& 4986 $\pm$ 111 & 		 &       & 		    & 3.525 $\pm$ 0.036 & $-$0.21  \\
$\beta$ Hyi	 	& 2151 			& 5840 $\pm$ 59  & 5873 $\pm$ 45 & 5870  & 		    & 3.955 $\pm$ 0.018 & $-$0.10  \\
$\beta$ Vir	 	& 102870 		& 6012 $\pm$ 64  & 6083 $\pm$ 41 & 6150  & 		    & 4.125 $\pm$ 0.018 & 0.12 \\
$\delta$ Eri 		& 23249 		& 4986 $\pm$ 57  & 4954 $\pm$ 30 &       & 		    & 3.827 $\pm$ 0.018 & 0.15 \\
$\delta$ Pav 		& 190248 		& 		 & 		 & 5540  & 		    & 4.306 $\pm$ 0.034 & 0.38 \\
$\eta$ Boo 		& 121370 		& 6028 $\pm$ 47  & 6099 $\pm$ 28 & 6025  & 		    & 3.822 $\pm$ 0.019 & 0.24 \\
$\eta$ Ser 		& 168723 		& 		 &		 &       & 		    & 3.029 $\pm$ 0.037 & $-$0.11 \\
$\gamma$ Pav 		& 203608 		& 		 & 		 & 6135  & 		    & 4.397 $\pm$ 0.022 & $-$0.74 \\
$\gamma$ Ser 		& 142860 		& 		 & 		 & 6245  & 		    & 4.169 $\pm$ 0.032 & $-$0.26 \\
HR 5803			& 139211 		& 		 & 		 & 6280  & 		    & 4.229 $\pm$ 0.023 & $-$0.04 \\
$\iota$ Hor	 	& 17051 		& 		 & 		 & 6110  & 		    & 4.399 $\pm$ 0.022 & 0.15 \\
$\xi$ Hya		& 100407 		& 4984 $\pm$ 54  & 5044 $\pm$ 40 &       & 		    & 2.883 $\pm$ 0.032 & 0.21 \\
$\mu$ Ara 		& 160691 		& 		 & 		 & 5690  & 		    & 4.228 $\pm$ 0.023 & 0.32 \\
$\nu$ Ind 		& 211998 		& 		 & 		 &       & 		    & 3.432 $\pm$ 0.035 & $-$1.63 \\
Procyon A 		& 61421 		& 6494 $\pm$ 48  & 6554 $\pm$ 84 & 6595  & 3.976 $\pm$ 0.016  & 3.972 $\pm$ 0.018 & 0.01  \\
$\tau$ Cet 		& 10700 		& 5383 $\pm$ 47  & 5414 $\pm$ 21 & 5420  & 		    & 4.533 $\pm$ 0.018 & $-$0.48 \\
$\tau$ PsA 		& 210302 		& 		 & 		 & 6385  & 		    & 4.240 $\pm$ 0.021 & 0.01 \\

\hline
	\end{tabular}
\end{minipage}
\end{table*} 

\begin{table*}
\centering
\begin{minipage}{130mm}
	\caption{The effective temperatures derived in this work. The second column gives the \teff\ based on the abundances from the laboratory list. The third column gives the constrained \teff\ after fixing the \logg\ to the asteroseismic value. The fourth and fifth columns give the \teff\ determined when using abundances calculated differentially to the Sun, with a solar \mictrb\ of 1 and 0.85 \kms\ respectively. }
	\label{tefftable}
	\begin{tabular}{llllll} 
		\hline
Star name 	& 	  Laboratory    & Fixing \logg\   & Differential & Differential \\ 
 & & &  \mictrb\ 1 \kms\ & \mictrb\ 0.85 \kms\ \\
\hline
171 Pup		& 	  5747 $\pm$ 37 & 5771 $\pm$ 38  & 5760 $\pm$ 35 & 5783 $\pm$ 36 \\
70 Oph A	& 	  5300 $\pm$ 43 & 5295 $\pm$ 55  & 5355 $\pm$ 23 & 5333 $\pm$ 23 \\
$\alpha$ Cen A 	&  	  5799 $\pm$ 38 & 5792 $\pm$ 37  & 5825 $\pm$ 21 & 5826 $\pm$ 21 \\
$\alpha$ Cen B 	& 	  5197 $\pm$ 52 & 5189 $\pm$ 82  & 5220 $\pm$ 31 & 5202 $\pm$ 31 \\
$\alpha$ For 	& 	  6281 $\pm$ 47 & 6275 $\pm$ 49  & 6253 $\pm$ 37 & 6289 $\pm$ 40 \\
$\alpha$ Men 	& 	  5606 $\pm$ 38 & 5614 $\pm$ 36  & 5627 $\pm$ 17 & 5620 $\pm$ 18 \\
$\beta$ Aql	& 	  5082 $\pm$ 41 & 5072 $\pm$ 50  & 5103 $\pm$ 24 & 5089 $\pm$ 26 \\
$\beta$ Hyi	& 	  5870 $\pm$ 37 & 5875 $\pm$ 36  & 5864 $\pm$ 25 & 5874 $\pm$ 26 \\
$\beta$ Vir	& 	  6224 $\pm$ 46 & 6231 $\pm$ 43  & 6209 $\pm$ 29 & 6232 $\pm$ 30 \\
$\delta$ Eri 	& 	  4976 $\pm$ 48 & 4960 $\pm$ 55  & 5023 $\pm$ 31 & 5005 $\pm$ 33 \\
$\delta$ Pav 	& 	  5576 $\pm$ 47 & 5560 $\pm$ 47  & 5611 $\pm$ 21 & 5599 $\pm$ 22 \\
$\eta$ Boo 	& 	  6205 $\pm$ 83 & 6231 $\pm$ 107 & 6200 $\pm$ 60 & 6214 $\pm$ 61 \\
$\eta$ Ser 	& 	  4888 $\pm$ 48 & 4880 $\pm$ 53  & 4911 $\pm$ 21 & 4894 $\pm$ 23 \\
$\gamma$ Pav 	& 	  6157 $\pm$ 51 & 6167 $\pm$ 51  & 6085 $\pm$ 39 & 6131 $\pm$ 42 \\
$\gamma$ Ser 	& 	  6350 $\pm$ 55 & 6368 $\pm$ 50  & 6321 $\pm$ 40 & 6363 $\pm$ 41 \\
HR 5803		& 	  6385 $\pm$ 47 & 6416 $\pm$ 46  & 6363 $\pm$ 35 & 6395 $\pm$ 37 \\
$\iota$ Hor	& 	  6215 $\pm$ 45 & 6205 $\pm$ 44  & 6218 $\pm$ 34 & 6236 $\pm$ 36 \\
$\xi$ Hya	& 	  5097 $\pm$ 45 & 5101 $\pm$ 51  & 5120 $\pm$ 29 & 5106 $\pm$ 29 \\
$\mu$ Ara 	& 	  5764 $\pm$ 39 & 5757 $\pm$ 34  & 5772 $\pm$ 22 & 5772 $\pm$ 29 \\
$\nu$ Ind 	& 	  5218 $\pm$ 36 & 5212 $\pm$ 37  & 5178 $\pm$ 31 & 5195 $\pm$ 33 \\
Procyon A 	& 	  6645 $\pm$ 46 & 6648 $\pm$ 42  & 6583 $\pm$ 38 & 6650 $\pm$ 38 \\
$\tau$ Cet 	& 	  5317 $\pm$ 41 & 5322 $\pm$ 42	 & 5320 $\pm$ 22 & 5308 $\pm$ 23 \\
$\tau$ PsA 	& 	  6494 $\pm$ 71 & 6511 $\pm$ 88  & 6426 $\pm$ 44 & 6473 $\pm$ 45 \\

\hline
	\end{tabular}
\end{minipage}
\end{table*}

\begin{table*}
\centering
\begin{minipage}{130mm}
	\caption{The \logg\ determined in this work. The second column gives the value based on the ionisation balance using the laboratory line list. The third and fourth  columns give the \logg\ determined when using abundances calculated differentially to the Sun, with a solar \mictrb\ of 1 and 0.85 \kms\ respectively. The last two columns give the \logg\ calculated from the pressure broadened Mg~{\sc i} b and Na~{\sc i} D lines.}
	\label{loggtable}
	\begin{tabular}{llllll} 
		\hline
		Star name &  Laboratory & Differential & Differential  & Mg~{\sc i} b & Na~{\sc i} D \\
		 & & \mictrb\ 1 \kms\ & \mictrb\ 0.85 \kms\ & & \\
				\hline
171 Pup			 & 4.17 $\pm$ 0.06  & 4.19 $\pm$ 0.06  & 4.22 $\pm$ 0.06  & 3.83  $\pm$  0.10 & 4.00 $\pm$	0.12 \\
70 Oph A		 & 4.35 $\pm$ 0.10  & 4.46 $\pm$ 0.06  & 4.43 $\pm$ 0.06  & 4.28  $\pm$  0.15 & 4.45 $\pm$	0.12 \\
$\alpha$ Cen A 		 & 4.25 $\pm$ 0.09  & 4.32 $\pm$ 0.04  & 4.32 $\pm$ 0.04  & 4.02  $\pm$  0.20 & 4.25 $\pm$	0.10 \\
$\alpha$ Cen B 		 & 4.36 $\pm$ 0.12  & 4.40 $\pm$ 0.07  & 4.38 $\pm$ 0.07  & 4.07  $\pm$  0.30 & 4.50 $\pm$	0.15 \\
$\alpha$ For 	 	 & 4.09 $\pm$ 0.09  & 4.03 $\pm$ 0.07  & 4.08 $\pm$ 0.07  & 4.10  $\pm$  0.40 & 4.10 $\pm$	0.10 \\
$\alpha$ Men 		 & 4.40 $\pm$ 0.09  & 4.44 $\pm$ 0.04  & 4.44 $\pm$ 0.04  & 4.30  $\pm$  0.10 & 4.45 $\pm$	0.13 \\
$\beta$ Aql	 	 & 3.44 $\pm$ 0.11  & 3.47 $\pm$ 0.07  & 3.45 $\pm$ 0.07  & 3.27  $\pm$  0.10 & 3.40 $\pm$	0.15 \\
$\beta$ Hyi	 	 & 3.95 $\pm$ 0.08  & 3.95 $\pm$ 0.04  & 3.96 $\pm$ 0.04  & 3.75  $\pm$  0.17 & 3.95 $\pm$	0.10 \\
$\beta$ Vir	 	 & 4.22 $\pm$ 0.08  & 4.17 $\pm$ 0.06  & 4.19 $\pm$ 0.06  & 4.07  $\pm$  0.15 & 4.30 $\pm$	0.20 \\
$\delta$ Eri 		 & 3.49 $\pm$ 0.14  & 3.63 $\pm$ 0.08  & 3.60 $\pm$ 0.08  & 3.13  $\pm$  0.05 & 3.45 $\pm$	0.25 \\
$\delta$ Pav 		 & 4.18 $\pm$ 0.11  & 4.23 $\pm$ 0.08  & 4.21 $\pm$ 0.08  & 3.85  $\pm$  0.05 & 4.15 $\pm$	0.15 \\
$\eta$ Boo 		 & 4.04 $\pm$ 0.19  & 4.04 $\pm$ 0.09  & 4.06 $\pm$ 0.09  & 3.93  $\pm$  0.10 & 4.10 $\pm$	0.20 \\
$\eta$ Ser 		 & 2.96 $\pm$ 0.15  & 3.02 $\pm$ 0.07  & 3.00 $\pm$ 0.07  & 2.48  $\pm$  0.35 & 2.75 $\pm$	0.15 \\
$\gamma$ Pav 		 & 4.31 $\pm$ 0.08  & 4.18 $\pm$ 0.08  & 4.24 $\pm$ 0.08  & 4.23  $\pm$  0.20 & 4.10 $\pm$	0.20 \\
$\gamma$ Ser 		 & 4.17 $\pm$ 0.09  & 4.18 $\pm$ 0.06  & 4.23 $\pm$ 0.06  & 4.13  $\pm$  0.15 & 4.33 $\pm$	0.05 \\
HR 5803			 & 4.21 $\pm$ 0.08  & 4.18 $\pm$ 0.06  & 4.22 $\pm$ 0.06  & 4.15  $\pm$  0.10 & 4.10 $\pm$	0.05 \\
$\iota$ Hor	 	 & 4.43 $\pm$ 0.08  & 4.43 $\pm$ 0.06  & 4.45 $\pm$ 0.06  & 4.25  $\pm$  0.15 & 4.35 $\pm$	0.05 \\
$\xi$ Hya		 & 2.93 $\pm$ 0.14  & 2.94 $\pm$ 0.07  & 2.92 $\pm$ 0.07  & 2.27  $\pm$  0.30 & 2.80 $\pm$	0.40 \\
$\mu$ Ara 		 & 4.20 $\pm$ 0.08  & 4.20 $\pm$ 0.05  & 4.21 $\pm$ 0.05  & 4.02  $\pm$  0.20 & 4.20 $\pm$	0.20 \\
$\nu$ Ind 		 & 3.40 $\pm$ 0.10  & 3.26 $\pm$ 0.09  & 3.29 $\pm$ 0.09  & 2.92  $\pm$  0.10 & 2.90 $\pm$	0.10 \\
Procyon A 		 & 3.96 $\pm$ 0.07  & 3.90 $\pm$ 0.05  & 3.93 $\pm$ 0.05  & 3.98  $\pm$  0.05 & 4.00 $\pm$	0.05 \\
$\tau$ Cet 		 & 4.49 $\pm$ 0.10    & 4.47 $\pm$ 0.05  & 4.46 $\pm$ 0.05  & 4.20  $\pm$  0.30 & 4.30 $\pm$	0.10 \\
$\tau$ PsA 		 & 4.31 $\pm$ 0.11  & 4.27 $\pm$ 0.07  & 4.32 $\pm$ 0.07  & 4.52  $\pm$  0.25 & 4.35 $\pm$	0.05 \\
\hline
	\end{tabular}
\end{minipage}
\end{table*}		

\begin{table*}
\centering
\begin{minipage}{130mm}
	\caption{The metallicity determined in this work. All abundances are relative to \citet{Asplund09}. The second column gives the [Fe/H] using the laboratory line list.  The third and fourth  columns give the [Fe/H] determined when using abundances calculated differentially to the Sun, with a solar \mictrb\ of 1 and 0.85 \kms\ respectively. The fifth and sixth columns give the [Mg/H] and [Na/H] abundances respectively. }
	\label{FeHtable}
	\begin{tabular}{lrrrrrr} 
		\hline
		Star name & Laboratory  & Differential & Differential & {[Mg/H]} & {[Na/H]} \\
		  & &  \mictrb\ 1 \kms\ & \mictrb\ 0.85 \kms\ & &  \\
		\hline
171 Pup			&  $-$0.76 $\pm$ 0.09 	& $-$0.82	 $\pm$ 0.08 & $-$0.82	 $\pm$ 0.08 & $-$0.59  $\pm$   0.07  & 	$-$0.59  $\pm$   0.08 \\
70 Oph A		&  0.14	 $\pm$ 0.09 	& 0.08	 $\pm$ 0.05 & 0.08	 $\pm$ 0.05 & 0.21   $\pm$   0.07  & 	0.18   $\pm$   0.07 \\
$\alpha$ Cen A  	&  0.31	 $\pm$ 0.08	& 0.24	 $\pm$ 0.05 & 0.25	 $\pm$ 0.05 & 0.39   $\pm$   0.10  & 	0.44   $\pm$   0.08 \\
$\alpha$ Cen B  	&  0.29	 $\pm$ 0.10 	& 0.21	 $\pm$ 0.06 & 0.21	 $\pm$ 0.07 & 0.51   $\pm$   0.12  & 	0.45   $\pm$   0.09 \\
$\alpha$ For 	        &  $-$0.09 $\pm$ 0.08 	& $-$0.18	 $\pm$ 0.07 & $-$0.17	 $\pm$ 0.07 & $-$0.16  $\pm$   0.21  & 	$-$0.09  $\pm$   0.05 \\
$\alpha$ Men 	   	&  0.21	 $\pm$ 0.09 	& 0.14	 $\pm$ 0.05 & 0.14	 $\pm$ 0.05 & 0.29   $\pm$   0.05  & 	0.28   $\pm$   0.09 \\
$\beta$ Aql	 	&  $-$0.09 $\pm$ 0.10 	& $-$0.14	 $\pm$ 0.07 & $-$0.15	 $\pm$ 0.07 & 0.06   $\pm$   0.07  & 	$-$0.04  $\pm$   0.08 \\
$\beta$ Hyi	 	&  0.01	 $\pm$ 0.08 	& $-$0.08	 $\pm$ 0.06 & $-$0.08	 $\pm$ 0.06 & 0.11   $\pm$   0.07  & 	0.06   $\pm$   0.09 \\
$\beta$ Vir	 	&  0.27	 $\pm$ 0.09	& 0.18	 $\pm$ 0.06 & 0.19	 $\pm$ 0.06 & 0.27   $\pm$   0.09  & 	0.29   $\pm$   0.11 \\
$\delta$ Eri 		&  0.15	 $\pm$ 0.11 	& 0.09	 $\pm$ 0.09 & 0.09	 $\pm$ 0.09 & 0.38   $\pm$   0.02  & 	0.30   $\pm$   0.13 \\
$\delta$ Pav 		&  0.43	 $\pm$ 0.09	& 0.37	 $\pm$ 0.06 & 0.37	 $\pm$ 0.06 & 0.59   $\pm$   0.02  & 	0.59   $\pm$   0.10 \\
$\eta$ Boo 		&  0.40	 $\pm$ 0.12	& 0.31	 $\pm$ 0.09 & 0.31	 $\pm$ 0.09 & 0.37    $\pm$ 0.04   & 	0.60   $\pm$   0.09 \\
$\eta$ Ser 		&  $-$0.11 $\pm$ 0.11 	& $-$0.17	 $\pm$ 0.05 & $-$0.18	 $\pm$ 0.05 & 0.06   $\pm$   0.14  & 	$-$0.05  $\pm$   0.09 \\
$\gamma$ Pav 		&  $-$0.58 $\pm$ 0.09 	& $-$0.69	 $\pm$ 0.08 & $-$0.69	 $\pm$ 0.08 & $-$0.50  $\pm$   0.13  & 	$-$0.38  $\pm$   0.13 \\
$\gamma$ Ser 		&  $-$0.12 $\pm$ 0.09 	& $-$0.22	 $\pm$ 0.06 & $-$0.21	 $\pm$ 0.06 & $-$0.10  $\pm$   0.06  & 	$-$0.06  $\pm$   0.01 \\
HR 5803			&  0.10	 $\pm$ 0.08 	& 0.01	 $\pm$ 0.06 & 0.02	 $\pm$ 0.06 & 0.08   $\pm$   0.08  & 	0.09   $\pm$   0.00 \\
$\iota$ Hor	 	&  0.23	 $\pm$ 0.09 	& 0.15	 $\pm$ 0.07 & 0.16	 $\pm$ 0.07 & 0.28   $\pm$   0.08  & 	0.28   $\pm$   0.05 \\
$\xi$ Hya		&  0.26	 $\pm$ 0.09 	& 0.18	 $\pm$ 0.07 & 0.18	 $\pm$ 0.07 & 0.43   $\pm$   0.10  & 	0.50   $\pm$   0.15 \\
$\mu$ Ara 		&  0.38	 $\pm$ 0.08 	& 0.30	 $\pm$ 0.05 & 0.30	 $\pm$ 0.05 & 0.43   $\pm$   0.09  & 	0.51   $\pm$   0.10 \\
$\nu$ Ind 		&  $-$1.46 $\pm$ 0.10 	& $-$1.56	 $\pm$ 0.09 & $-$1.57	 $\pm$ 0.09 & $-$1.09  $\pm$   0.05  & 	$-$1.54   $\pm$ 0.05	\\
Procyon A 		&  0.04	 $\pm$ 0.08	& -0.06	 $\pm$ 0.06 & -0.04	 $\pm$ 0.06 & 0.05   $\pm$   0.07  & 	0.10   $\pm$   0.02 \\
$\tau$ Cet 		&  $-$0.47 $\pm$ 0.10 	& $-$0.55  $\pm$ 0.09 & -0.56	 $\pm$ 0.10 & $-$0.11  $\pm$   0.07  & 	$-$0.33  $\pm$   0.06 \\
$\tau$ PsA 		&  0.18	 $\pm$ 0.10 	& 0.05	 $\pm$ 0.06 & 0.06	 $\pm$ 0.06 & $-$0.03  $\pm$   0.16  & 	0.12   $\pm$   0.03 \\
		                                
\hline
	\end{tabular}
	\end{minipage}
\end{table*} 	

\begin{table}
\centering
	\caption{The \mictrb\ determined in this work. The errors are 0.03 \kms. The second column gives the \mictrb\ based on the laboratory line list. The third and fourth columns give the \mictrb\ when using abundances calculated differentially to the Sun, with a solar \mictrb\ of 1 and 0.85 \kms\ respectively.}
	\label{micttable}
	\begin{tabular}{lrrrr} 
		\hline
		Star name & Laboratory  & Differential & Differential \\
		 & & 1 \kms\ & 0.85 \kms\ \\
				\hline		
171 Pup		 & 0.81   & 1.15  & 1.02 \\
70 Oph A	 & 0.79   & 0.94  & 0.79 \\
$\alpha$ Cen A   & 0.87   & 1.02  & 0.89 \\
$\alpha$ Cen B   & 0.74   & 0.85  & 0.71 \\
$\alpha$ For 	 & 1.19   & 1.37  & 1.28 \\
$\alpha$ Men 	 & 0.74   & 0.96  & 0.80 \\
$\beta$ Aql	 & 0.91   & 0.96  & 0.86 \\
$\beta$ Hyi	 & 1.00   & 1.17  & 1.07 \\
$\beta$ Vir	 & 1.12   & 1.29  & 1.18 \\
$\delta$ Eri 	 & 0.86   & 0.94  & 0.85 \\
$\delta$ Pav 	 & 0.80   & 0.94  & 0.80 \\
$\eta$ Boo 	 & 1.36   & 1.49  & 1.41 \\
$\eta$ Ser 	 & 1.02   & 1.09  & 1.01 \\
$\gamma$ Pav 	 & 1.00   & 1.29  & 1.19 \\
$\gamma$ Ser 	 & 1.34   & 1.54  & 1.45 \\
HR 5803		 & 1.17   & 1.34  & 1.25 \\
$\iota$ Hor	 & 0.96   & 1.14  & 1.02 \\
$\xi$ Hya	 & 1.18   & 1.25  & 1.18 \\
$\mu$ Ara 	 & 0.90   & 1.07  & 0.95 \\
$\nu$ Ind 	 & 1.06   & 1.31  & 1.23 \\
Procyon A 	 & 1.61   & 1.80  & 1.72 \\
$\tau$ Cet 	 &    & 0.62  & 0.36 \\
$\tau$ PsA	 & 1.23   & 1.47  & 1.39 \\	
\hline
	\end{tabular}
\end{table} 		
				
In this section, we compare the results from the laboratory and differential line lists to each other, and also compare the results from each list to the external parameters, i.e. the bolometric \teff, the photometric \teff, the binary \logg, and the asteroseismic \logg.			

\subsubsection{Temperature}
There is good agreement in \teff\ between the laboratory and differential lists. There is a difference of 2 $\pm$ 35 K between the laboratory and differential(1) list, and a difference of 9 $\pm$ 16 K between the laboratory and differential(0.85) list. 

\begin{figure}
	\includegraphics[height=\columnwidth,angle=270]{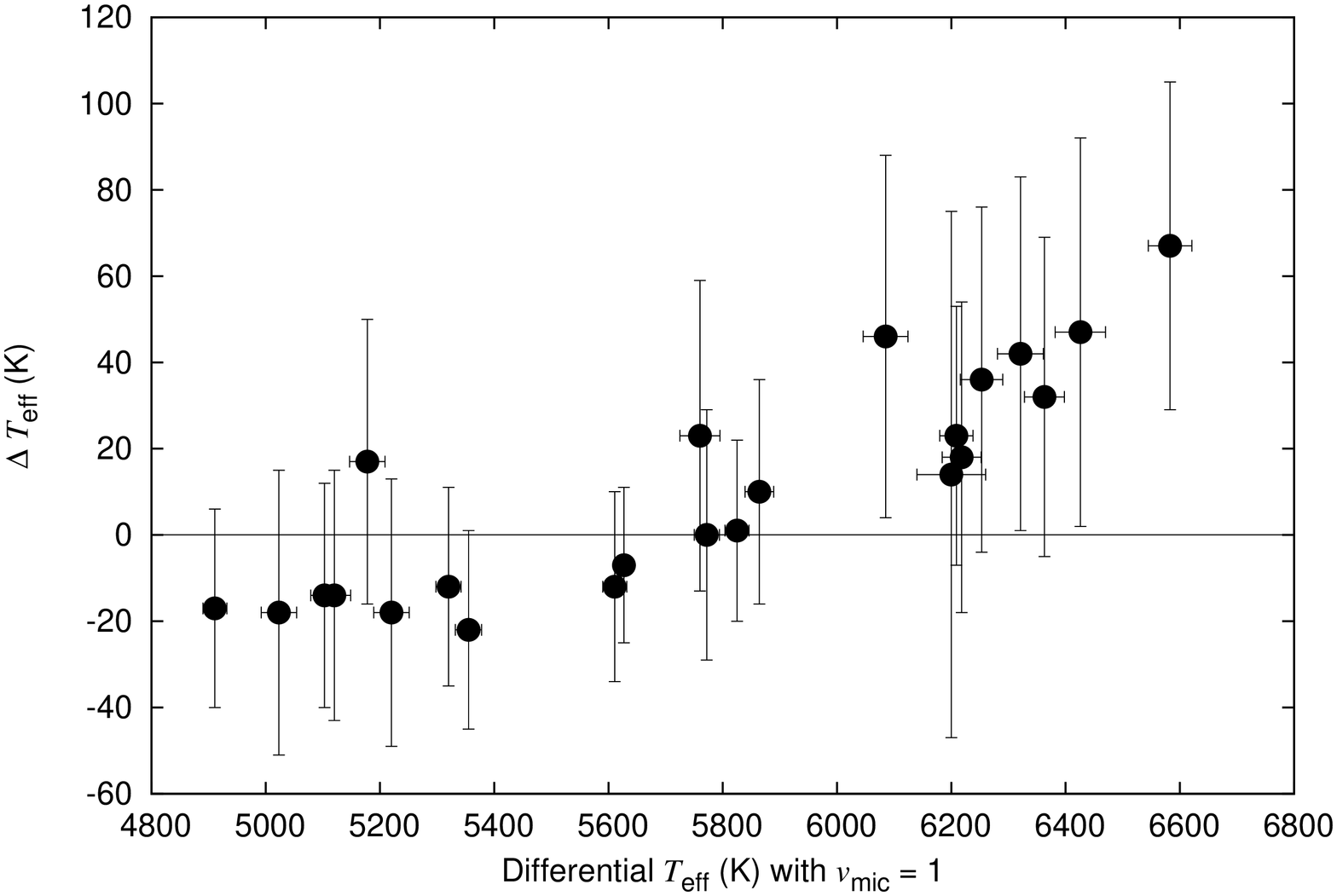}
    \caption{$\Delta$\teff\ = \teff\ differential(0.85) $-$ \teff\ differential(1). The \teff\ derived differentially with the solar \mictrb\ of 0.85 \kms\ is hotter for stars > 6000 K compared to the differential \teff\ derived with solar \mictrb\ of 1 \kms. }
    \label{teff_differential_mict}
\end{figure}

The difference in \teff\ between the two different differential lists is 11 $\pm$ 26 K. The \teff\ derived with the solar \mictrb\ of 0.85 \kms\ has a slightly higher \teff\ of 36 $\pm$ 17 K for stars hotter than 6000 K, as seen in Figure~\ref{teff_differential_mict}. The most noticeable difference is Procyon, where the \teff\ is 67 K higher. The difference in \teff\ is because \teff\ is dependent on \mictrb, and the output \mictrb\ for each star is dependent on which solar value was used initially to calculate the differential \loggf.

\begin{figure}
	\includegraphics[height=\columnwidth,angle=270]{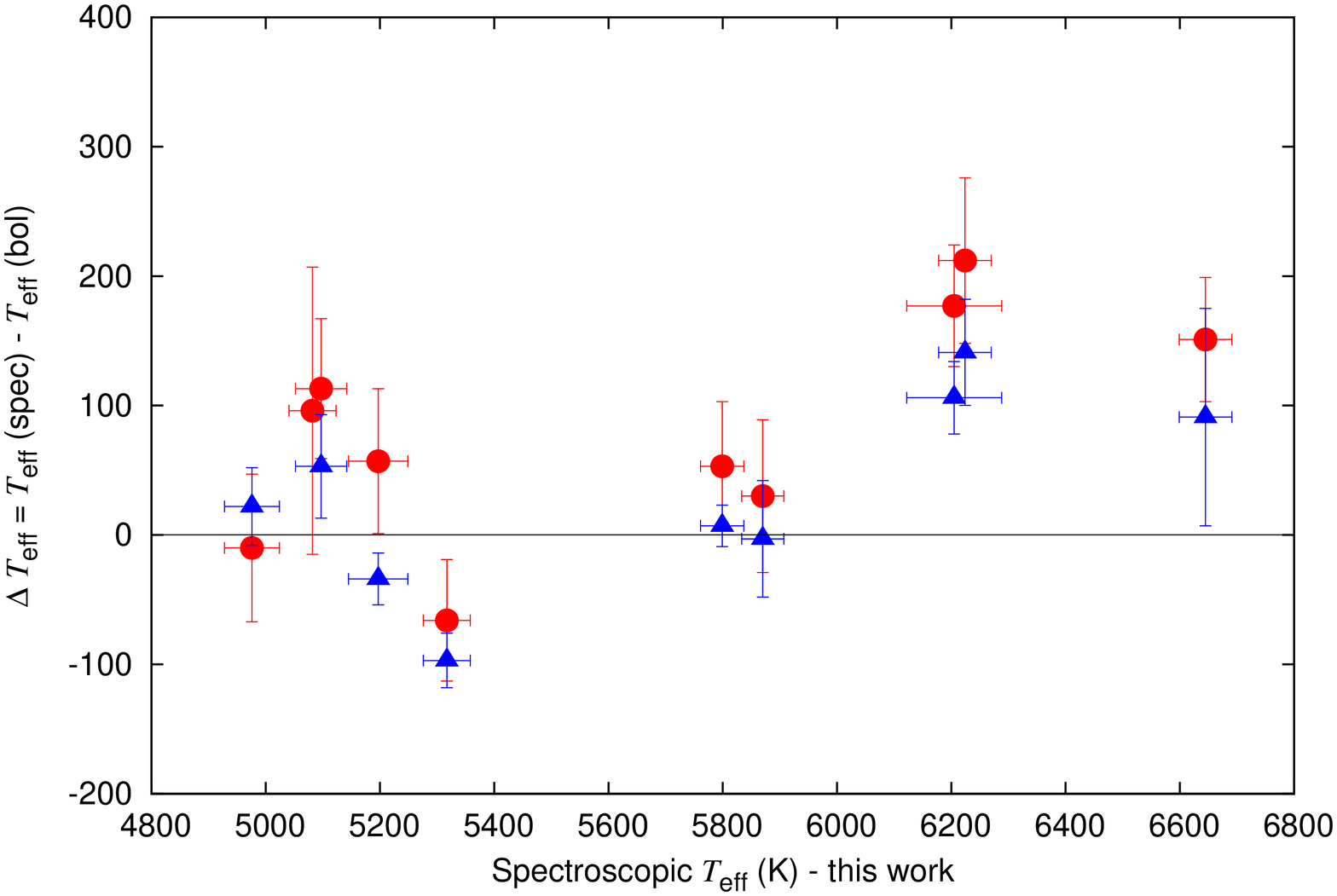}
	\includegraphics[height=\columnwidth,angle=270]{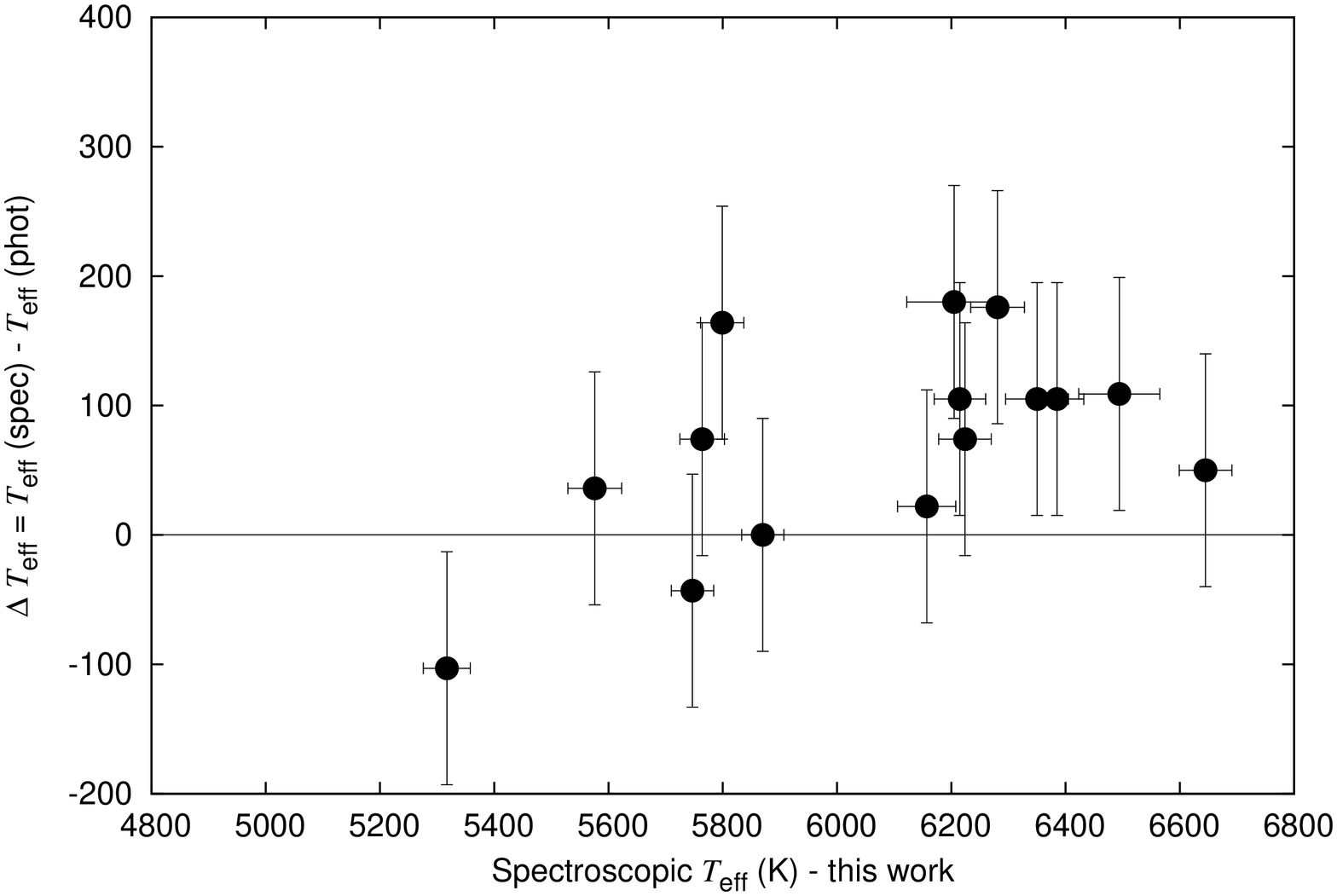}
	   \caption{The top panel shows the comparison between the spectroscopic \teff\ with the laboratory line list from this work and the bolometric \teff\ of \citet{Bruntt10} (red circles) and \citet{Heiter15} (blue triangles). The bottom panel shows the comparison between the spectroscopic \teff\ and the photometric \teff\ from B10. Our temperatures are hotter for stars with \teff\ > 6000 K.}
    \label{teffcomp}
\end{figure}

The bolometric \teff\ is a ``direct'' temperature determination in that it is almost independent of stellar models. The temperature is found from the bolometric flux and the angular diameter. The comparison between our spectroscopic \teff\ using the laboratory line list and the bolometric \teff\ is shown in the top panel of Figure~\ref{teffcomp}. The values from both B10 and H15 are shown. H15 compiled bolometric temperatures for their sample of 34 \textit{Gaia} FGK benchmark stars. The fact that there are two different versions of the bolometric \teff, which is a direct measurement and is supposed to represent the fundamental \teff\ of the star, is a matter of concern. It is difficult to compare our indirect values with the fundamental \teff\ when even those values are not certain. 

Our \teff\ using the laboratory list is hotter on average than B10 by 81 $\pm$ 86 K, and H15 by 32 $\pm$ 74 K. There is good agreement for the stars < 6000 K, particularly for the H15 values. For the three stars hotter than 6000 K, our spectroscopic \teff\ values using the laboratory list are significantly hotter, bringing the difference to 180 $\pm$ 30 K for B10 and 113 $\pm$ 26 K for H15. 

The differential(1) \teff\ is hotter than B10 by 87 $\pm$ 76 K and H15 by 36 $\pm$ 67 K. The differential(0.85) \teff\ is hotter than B10 by 91 $\pm$ 87  K and H15 by 41 $\pm$ 78 K. Therefore there is no advantage or disadvantage to using the laboratory line list for \teff. 

The comparison between our spectroscopic \teff\ using the laboratory list and the photometric \teff\ from B10, which is derived from Str{\"o}mgren photometric indices, is shown in the middle panel of Figure~\ref{teffcomp}. Our temperatures are systematically higher by 70 $\pm$ 80 K. The differential(1) \teff\ is hotter than the photometric \teff\ by 55 $\pm$ 80 K and the differential(0.85) \teff\  is hotter by 76 $\pm$ 81, K suggesting that the differential(1) list offers a slight improvement with \teff\ over the other two lists.

\subsubsection{Surface gravity}
\begin{figure}
	\includegraphics[height=\columnwidth,angle=270]{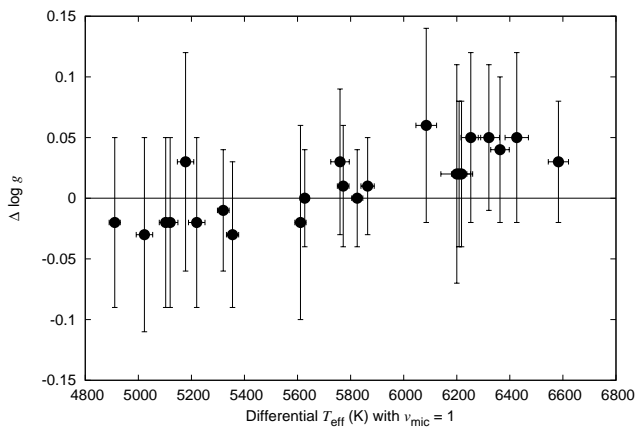}
    \caption{$\Delta$\logg\ = \logg\ differential(0.85) $-$ \logg\ differential(1). The \logg\ derived differentially with the solar \mictrb\ of 0.85 \kms\ is higher for stars > 6000 K compared to the differential \logg\ derived with solar \mictrb\ of 1 \kms.} 
    \label{delta_logg}
\end{figure}

There is good agreement between the laboratory and differential line lists for \logg. There is a difference of 0.00 $\pm$ 0.07 dex between laboratory and differential(1) and 0.01 $\pm$ 0.05 dex between laboratory and differential(0.85). The difference between the two differential lists is 0.01 $\pm$ 0.03 dex, however this depends on \teff\, as seen in Figure~\ref{delta_logg}.

Only four stars in this sample have \logg\ known from their binary nature. There is excellent agreement between the binary \logg\ for Procyon (3.976 $\pm$ 0.016 dex) our spectroscopic \logg\ using the laboratory list (3.96 $\pm$ 0.07 dex). The differential(1) and differential(0.85) lists give slightly lower \logg\ values, 3.90 $\pm$ 0.05 and 3.93 $\pm$ 0.05 dex respectively.

The binary \logg\ for $\alpha$ Cen A is 4.307 $\pm$ 0.005 dex, which agrees with the spectroscopic \logg\ with the laboratory list of 4.25 $\pm$ 0.09 dex within the errors. The differential(1) and differential(0.85) values, which are both 4.32 $\pm$ 0.04 dex, are in much better agreement with the fundamental \logg. The same is true for 70 Oph A, where the spectroscopic \logg\ with the laboratory list (4.35 $\pm$ 0.10 dex) agrees with the binary \logg\ (4.468 $\pm$ 0.030) within the errors, but the differential(1) and differential(0.85) values of 4.46 $\pm$ 0.06 and 4.43 $\pm$ 0.06 dex respectively are in better agreement with the fundamental value.

The spectroscopic \logg\ with the laboratory list of 4.36 $\pm$ 0.12 dex does not agree with the binary \logg\ for $\alpha$ Cen B, which is 4.538 $\pm$ 0.008 dex. The differential(1) and differential (0.85) values of 4.40 $\pm$ 0.07 and 4.38 $\pm$ 0.07 dex respectively agree within the errors.

\begin{figure}
	\includegraphics[height=\columnwidth,angle=270]{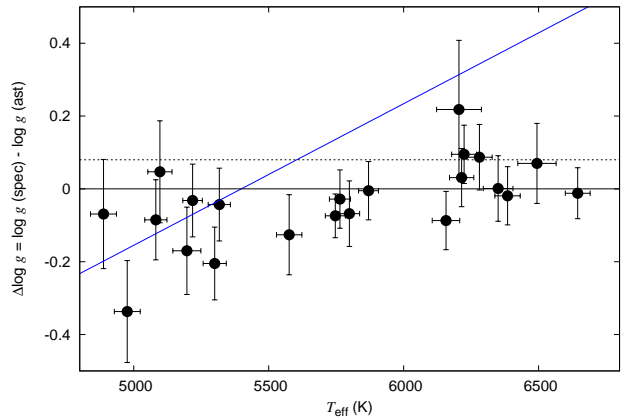}
    \caption{The difference between spectroscopic and asteroseismic \logg\ as a function of \teff. The dotted line at 0.08 dex is the mean difference between $\logg_{\rm spec}$ and $\logg_{\rm ast}$ found by \citet{Bruntt12}. The blue line is the linear trend seen in \citet{Mortier14}. There appears to be a \teff\ dependence in our data, but if the two outliers ($\delta$ Eri and $\eta$ Boo) are removed then this no longer holds. }
    \label{logg_diff_m}
\end{figure}

When comparing the spectroscopic \logg\ with the laboratory list to the asteroseismic \logg, there is a lot of dispersion as seen in Figure~\ref{logg_diff_m}. This figure shows the difference between the spectroscopic \logg\ that we derived and the asteroseismic \logg\ as a function of the spectroscopic \teff\ with the laboratory list. \citet{Bruntt12} found the mean difference between $\logg_{\rm spec}$ and $\logg_{\rm ast}$ to be 0.08 $\pm$ 0.07 dex for their sample of 93 solar-like \textit{Kepler} stars, shown as a dashed line. In an analysis of 76 planet host stars by \citet{Mortier14}, the difference was found to be dependent on \teff, and this is shown with the blue line.

The mean difference between the spectroscopic \logg\ with the laboratory list and the asteroseismic \logg\ is 0.04 $\pm$ 0.11 dex. There is no real improvement when using the differential lists, with the differential(1) list giving a mean difference of 0.04 $\pm$ 0.10 dex when compared to the asteroseismic \logg, and the differential(0.85) list giving a difference of 0.03 $\pm$ 0.10 dex.

\subsubsection{Metallicity}
The difference between [Fe/H] of the two differential analyses for our sample of stars is negligible; 0.00 $\pm$ 0.01 dex, showing that the choice of initial \mictrb\ does not affect the resulting abundance. 

\begin{figure}
	\includegraphics[height=\columnwidth,angle=270]{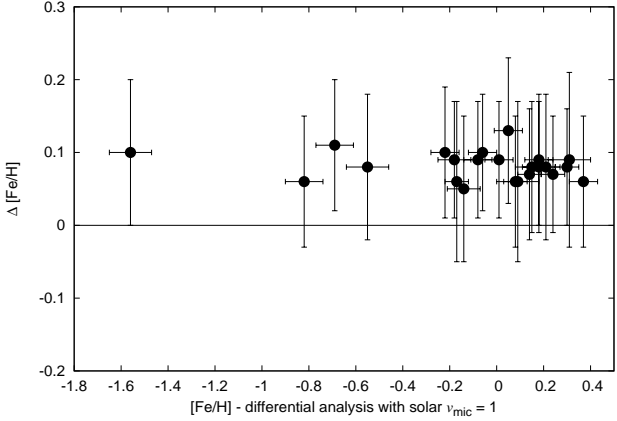}
    \caption{The comparison between the laboratory [Fe/H] and the differential [Fe/H] (assuming a solar \mictrb\ of 1 \kms). $\Delta$[Fe/H] = laboratory $-$ differential. The laboratory values are systematically higher.}
    \label{finalfe_differential}
\end{figure}

Our laboratory [Fe/H] is also systematically higher than both differential lists by 0.08 $\pm$ 0.02 dex, as seen in Figure~\ref{finalfe_differential}.

The laboratory [Fe/H] is also systematically higher than the B10 [Fe/H] by 0.09 $\pm$ 0.06 dex. Both differential lists are in good agreement with the B10 values (0.01 $\pm$ 0.05 dex for differential(1) and 0.01 $\pm$ 0.06 dex for differential(0.85)), which is to be expected as the B10 line list was also created differentially to the Sun.

\subsubsection{Microturbulence}
Unlike \teff\ and \logg\ which are physical parameters, the \mictrb\ of the Sun is model-dependent and is also strongly dependent on the line list used. While either solar \mictrb\ can be used when creating a differential line list, it is important to remember that using \loggf\ values calculated from the Sun will create a significant bias when trying to determine the \mictrb\ for another star. This is seen in Table~\ref{micttable}, where the differential(0.85) line list always results in a lower \mictrb\ for other stars compared to the differential(1) list, as expected. The systematic offset is 0.11 $\pm$ 0.04 \kms.

\subsection{Surface gravity from pressure broadened lines}

We also obtained the \logg\ for the pressure broadened Mg~{\sc i} b and Na~{\sc i} D lines, which are listed in Table~\ref{loggtable}. It is important to know the abundance of these elements before measuring the \logg, so the EWs of several weaker Mg~{\sc i} and Na~{\sc i} lines were measured in order to determine the abundance. These are given in Table~\ref{FeHtable}. The error for the pressure broadened \logg\ is determined from the error in the Na and Mg abundances.

The continuum is very difficult to measure around the Mg triplet, particularly for cooler stars with high metallicity. As such, the \logg\ from the Na~{\sc i} D lines should be prioritised, although it should be noted that these lines can be affected by interstellar absorption.  The Ca lines used in B10 were not used here as these lines are much more sensitive to changes in other broadening factors such as macroturbulence and rotational velocity compared to the Mg~{\sc i} b and Na~{\sc i} D lines. 

The \logg\ from the Mg triplet is significantly underestimated compared to the binary \logg, except for Procyon. The \logg\ from the Na~{\sc i} D lines on the other hand, agrees well with all of the binary \logg\ values.

The Mg \logg\ does not compare well with the asteroseismic \logg, giving a mean difference of 0.24 $\pm$ 0.25 dex. The Na \logg\ is somewhat better, with a mean difference of 0.09 $\pm$ 0.19 dex compared to the asteroseismic \logg.

There is no advantage to using the Mg triplet in order to get the spectroscopic \logg, as the values are highly discrepant from the binary and asteroseismic \logg. The Na~{\sc i} D returns reasonable results and can be used as an additional check of the spectroscopic \logg, but it does not offer any improvement over the \logg\ from the ionisation balance.

\subsection{Mass and radius}
The mass and radius were determined using the \citep{Torres10} calibration, which determines $M$ and $R$ from the \teff, \logg, and [Fe/H]. Two determinations of mass and radius are given in Table~\ref{massradius}. The first uses the spectroscopic parameters determined using the differential(1) list. The second uses the asteroseismic \logg, but the same values of \teff\ and [Fe/H] as in the first determination. There is an improvement in the precision for the radius.

\begin{table*}
\centering
\begin{minipage}{100mm}
	\caption{The mass and radius are given in units of $M_{\sun}$ and $R_{\sun}$. $M_{\rm spec}$  and $R_{\rm spec}$ are the values determined using spectroscopy with the differential(1) list. $M_{\rm ast}$  and $R_{\rm ast}$ use the asteroseismic \logg, but the same spectroscopic \teff\ and [Fe/H]. }
	\label{massradius}
	\begin{tabular}{lllll} 
		\hline
Star name & $M_{\rm spec}$  & $R_{\rm spec}$  & $M_{\rm ast}$  & $R_{\rm ast}$  \\   
 & differential(1) & differential(1) &  &  \\ \hline
171 Pup	  & 0.93 $\pm$  0.06  &  1.29 $\pm$ 0.11  & 0.91 $\pm$ 	0.06  & 1.20 $\pm$ 0.05 \\
70 Oph A    & 0.95 $\pm$	0.06  &  0.94 $\pm$ 0.08  & 0.93 $\pm$ 	0.06  & 0.84 $\pm$ 0.04 \\
$\alpha$ Cen A & 1.16 $\pm$	0.07  &  1.22 $\pm$ 0.07  & 1.16 $\pm$ 	0.07  & 1.22 $\pm$ 0.05 \\
$\alpha$ Cen B & 0.95 $\pm$	0.06  &  1.00 $\pm$ 0.09  & 0.93 $\pm$ 	0.06  & 0.85 $\pm$ 0.03 \\
$\alpha$ For  & 1.30 $\pm$	0.09  &  1.82 $\pm$ 0.19  & 1.31 $\pm$ 	0.09  & 1.89 $\pm$ 0.11 \\
$\alpha$ Men  & 1.04 $\pm$	0.07  &  1.01 $\pm$ 0.06  & 	              & 	        \\
$\beta$ Aql   & 1.31 $\pm$	0.10  &  3.50 $\pm$ 0.39  & 1.27 $\pm$ 	0.09  & 3.23 $\pm$ 0.20 \\
$\beta$ Hyi   & 1.25 $\pm$	0.08  &  1.96 $\pm$ 0.13  & 1.25 $\pm$ 	0.08  & 1.94 $\pm$ 0.08 \\
$\beta$ Vir   & 1.32 $\pm$	0.09  &  1.55 $\pm$ 0.13  & 1.34 $\pm$ 	0.08  & 1.64 $\pm$ 0.07 \\
$\delta$ Eri  & 1.22 $\pm$	0.10  &  2.78 $\pm$ 0.34  & 1.10 $\pm$ 	0.08  & 2.09 $\pm$ 0.09 \\
$\delta$ Pav  & 1.16 $\pm$	0.08  &  1.34 $\pm$ 0.15  & 1.14 $\pm$ 	0.08  & 1.22 $\pm$ 0.07 \\
$\eta$ Boo    & 1.43 $\pm$	0.11  &  1.87 $\pm$ 0.24  & 1.59 $\pm$ 	0.09  & 2.54 $\pm$ 0.11 \\
$\eta$ Ser    & 1.67 $\pm$	0.14  &  6.70 $\pm$ 0.75  & 1.66 $\pm$ 	0.12  & 6.61 $\pm$ 0.43 \\
$\gamma$ Pav  & 1.04 $\pm$	0.07  &  1.38 $\pm$ 0.16  & 0.98 $\pm$ 	0.07  & 1.04 $\pm$ 0.05 \\
$\gamma$ Ser  & 1.23 $\pm$	0.08  &  1.49 $\pm$ 0.13  & 1.23 $\pm$ 	0.08  & 1.51 $\pm$ 0.08 \\
HR 5803    & 1.31 $\pm$	0.09  &  1.53 $\pm$ 0.13  & 1.29 $\pm$ 	0.08  & 1.43 $\pm$ 0.06 \\
$\iota$ Hor   & 1.21 $\pm$	0.08  &  1.10 $\pm$ 0.09  & 1.22 $\pm$ 	0.08  & 1.15 $\pm$ 0.05 \\
$\xi$ Hya    & 2.06 $\pm$	0.17  &  8.11 $\pm$ 0.91  & 2.15 $\pm$ 	0.14  & 8.85 $\pm$ 0.52 \\
$\mu$ Ara     & 1.21 $\pm$	0.08  &  1.42 $\pm$ 0.11  & 1.20 $\pm$ 	0.08  & 1.37 $\pm$ 0.06 \\
$\nu$ Ind     & 1.11 $\pm$	0.10  &  4.25 $\pm$ 0.60  & 0.99 $\pm$ 	0.08  & 3.28 $\pm$ 0.20 \\ 
Procyon   & 1.53 $\pm$	0.10  &  2.30 $\pm$ 0.18  & 1.47 $\pm$ 	0.10  & 2.07 $\pm$ 0.09 \\
$\tau$ Cet    & 0.81 $\pm$	0.05  &  0.87 $\pm$ 0.06  & 0.80 $\pm$ 	0.05  & 0.80 $\pm$ 0.03 \\
$\tau$ PsA    & 1.30 $\pm$	0.09  &  1.37 $\pm$ 0.13  & 1.31 $\pm$ 	0.08  & 1.43 $\pm$ 0.06 \\
\hline
	\end{tabular}
	\end{minipage}
\end{table*}

\subsection{Line selection}
\label{lineselection}
\begin{figure}
	\includegraphics[height=\columnwidth,angle=270]{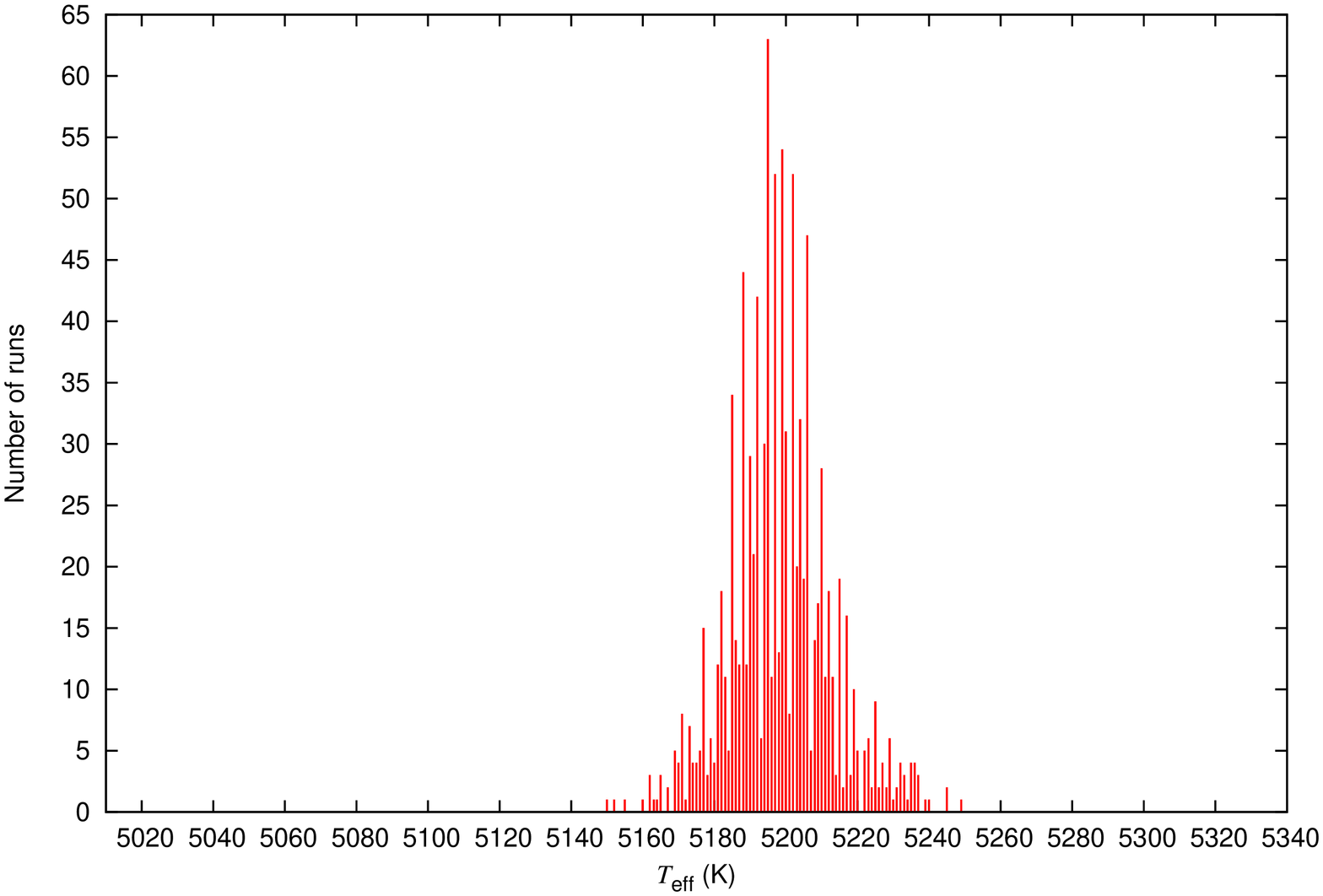}
	\includegraphics[height=\columnwidth,angle=270]{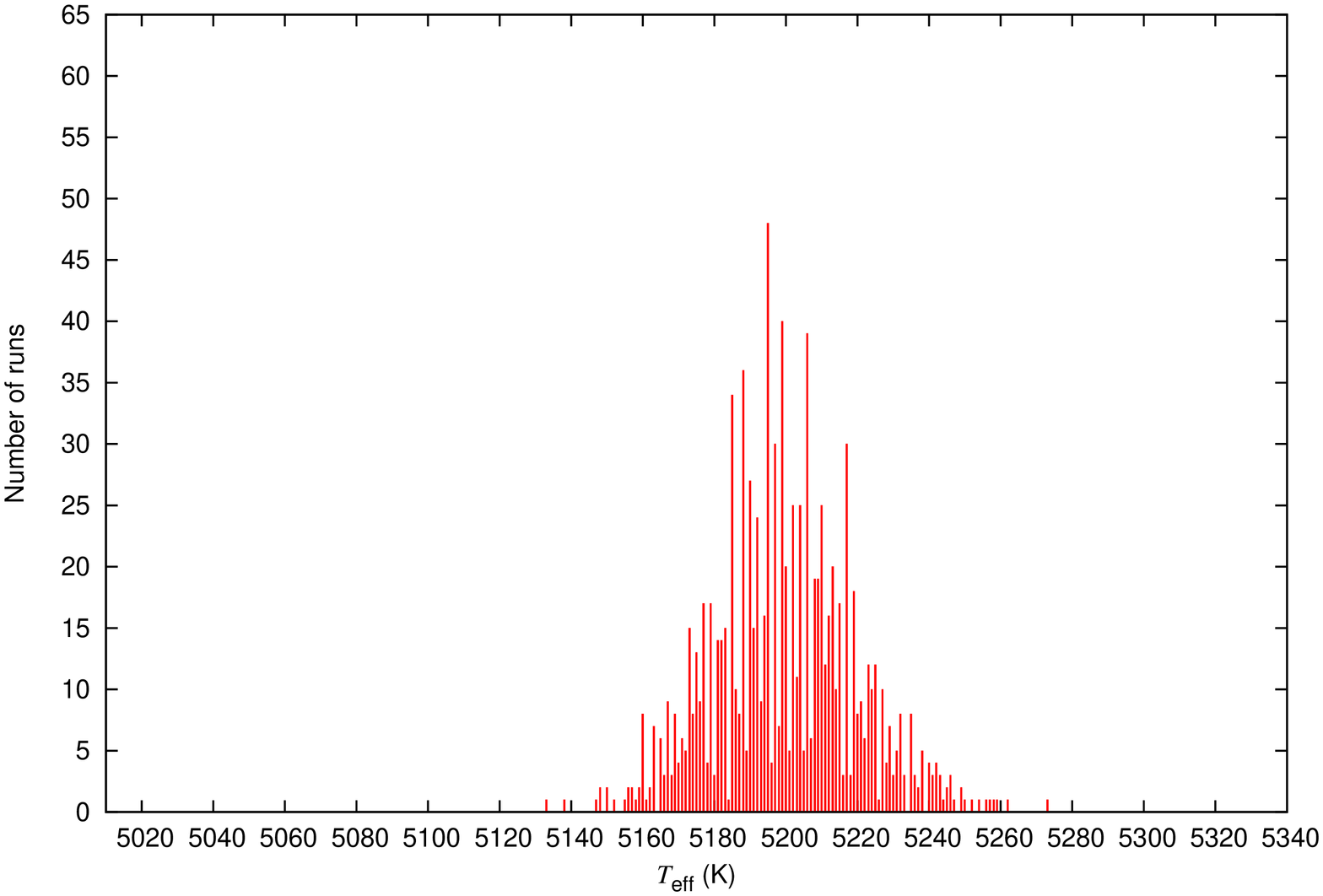}
	\includegraphics[height=\columnwidth,angle=270]{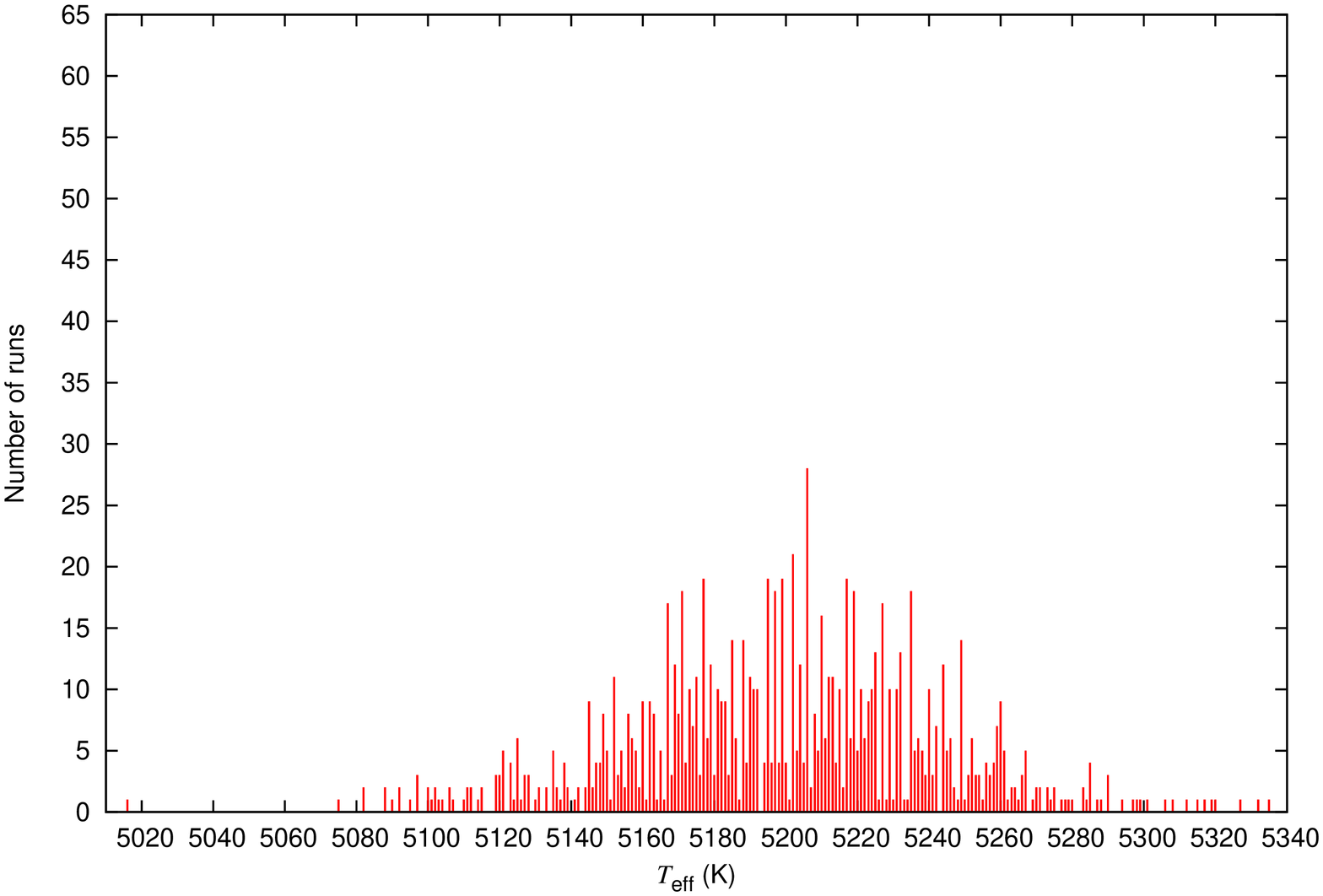}
    \caption{The \teff\ distribution for 1000 temperature runs of $\alpha$ Cen B with random Fe~{\sc i} lines removed. The top panel has 5 Fe~{\sc i} lines removed, the middle panel has 10 Fe~{\sc i} lines removed, and the bottom panel has 30 Fe~{\sc i} lines removed. The majority of runs still return the input value of 5197 K, but the scatter obviously increases when more lines are removed. }
    \label{alphacenB_FeI}
\end{figure}

The \teff\ derived from the excitation balance, and thus the \logg\ and [Fe/H], is sensitive to the exact lines used. $\alpha$ Cen B was used a test case and 1000 temperature runs were performed with 5 random Fe~{\sc i} lines removed for each run. The input parameters were those derived using the full set of lines with laboratory data; \teff\ = 5197 K, \logg\ = 4.34 dex and \mictrb\ = 0.74 \kms. The resulting \teff\ distribution is shown in the top panel of Figure~\ref{alphacenB_FeI}. The peak of the distribution is at 5198 $\pm$ 14 K. The middle panel shows the distribution with 10 Fe~{\sc i} lines removed, which has a peak at 5200 $\pm$ 20 K. The bottom panel has 30 Fe~{\sc i} lines removed with a peak at 5199 $\pm$ 42 K. With more lines removed, the scatter obviously increases and the result is that fewer runs return the input \teff. In low S/N ($\sim$50) CORALIE or SOPHIE spectra that are frequently used to characterise planet host stars, it is quite common for $\sim$30 lines to be unmeasurable when using the \citet{Doyle13} line list of 72 Fe lines.

\begin{figure}
	\includegraphics[height=\columnwidth,angle=270]{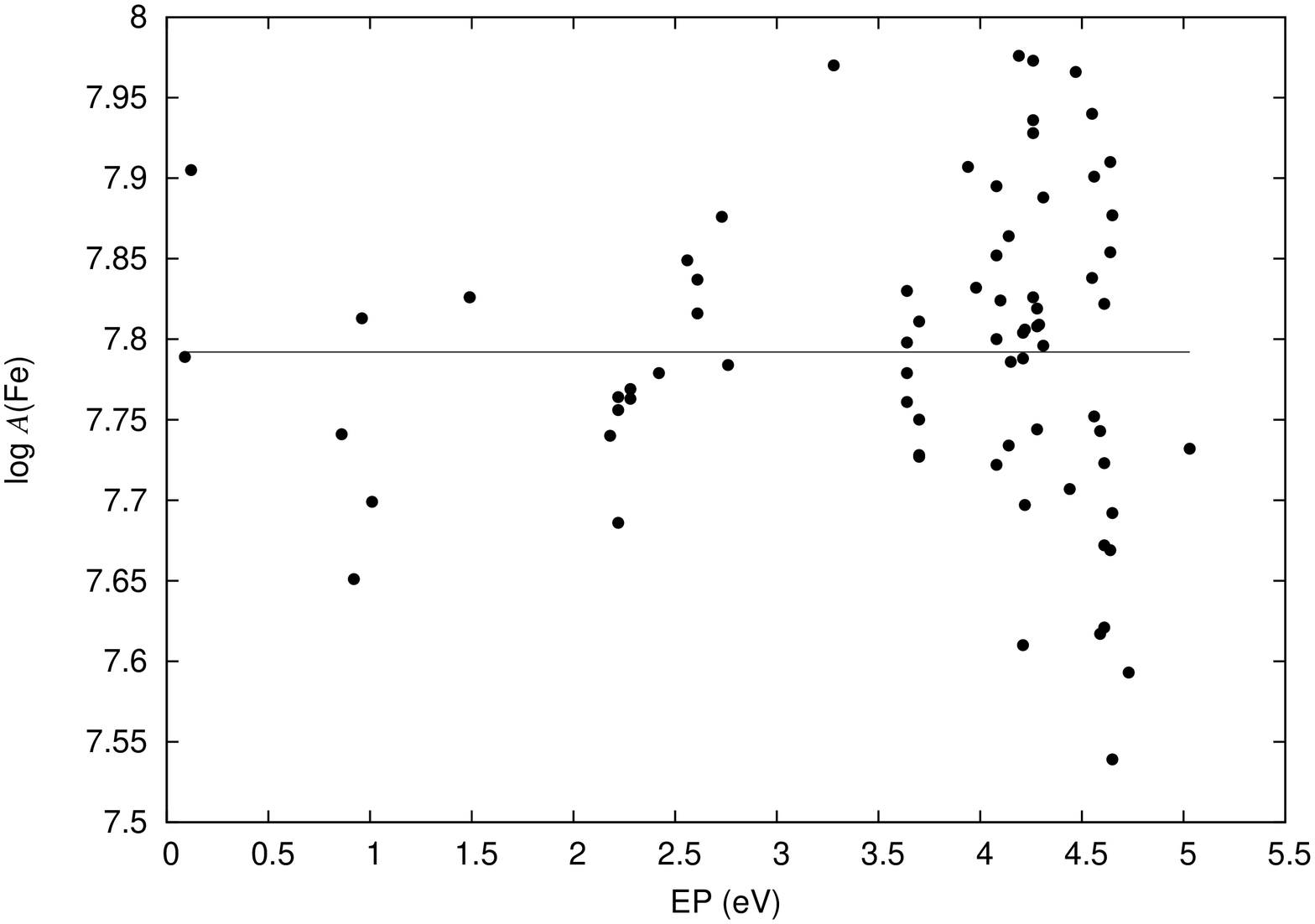}
	\includegraphics[height=\columnwidth,angle=270]{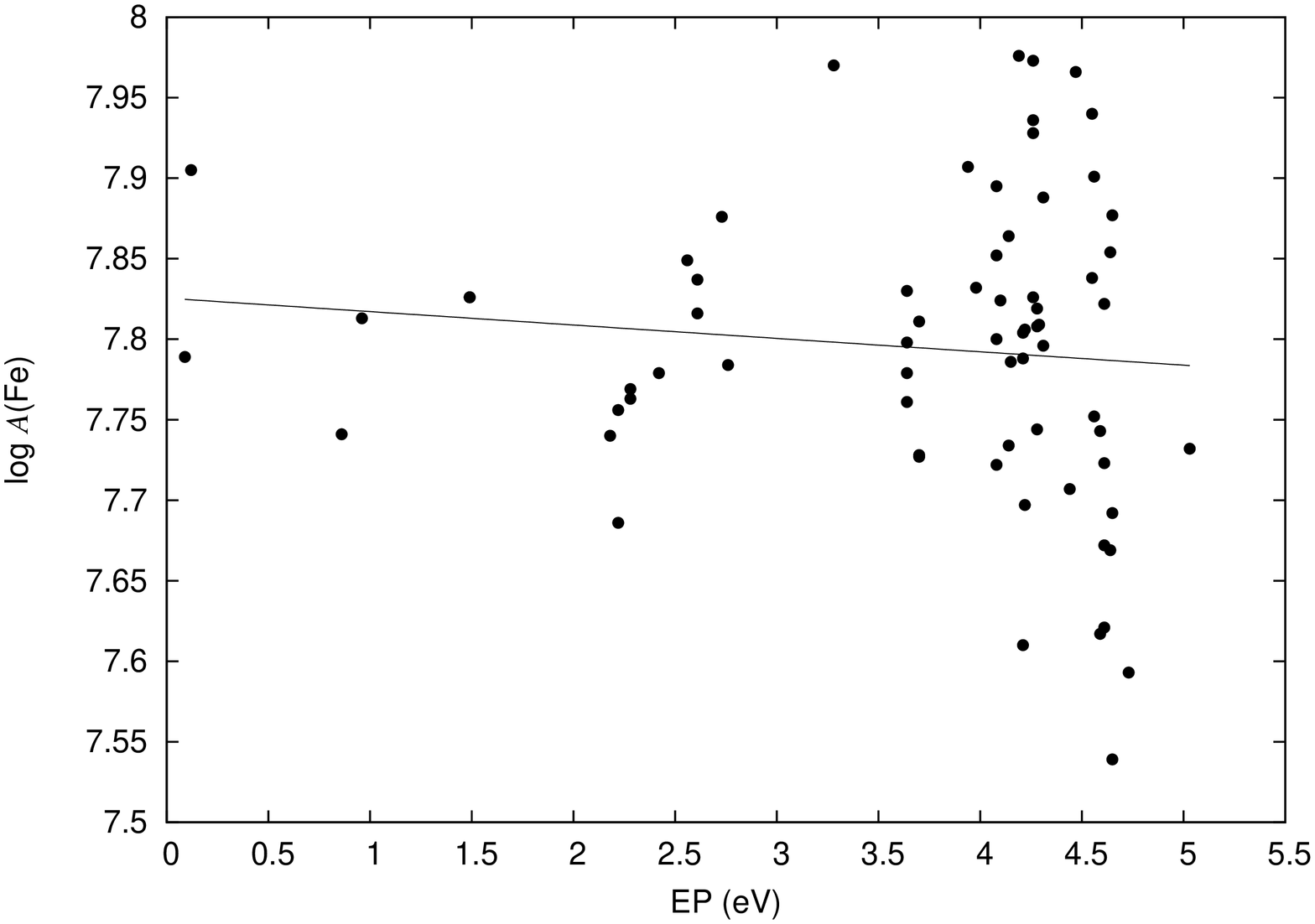}
	\includegraphics[height=\columnwidth,angle=270]{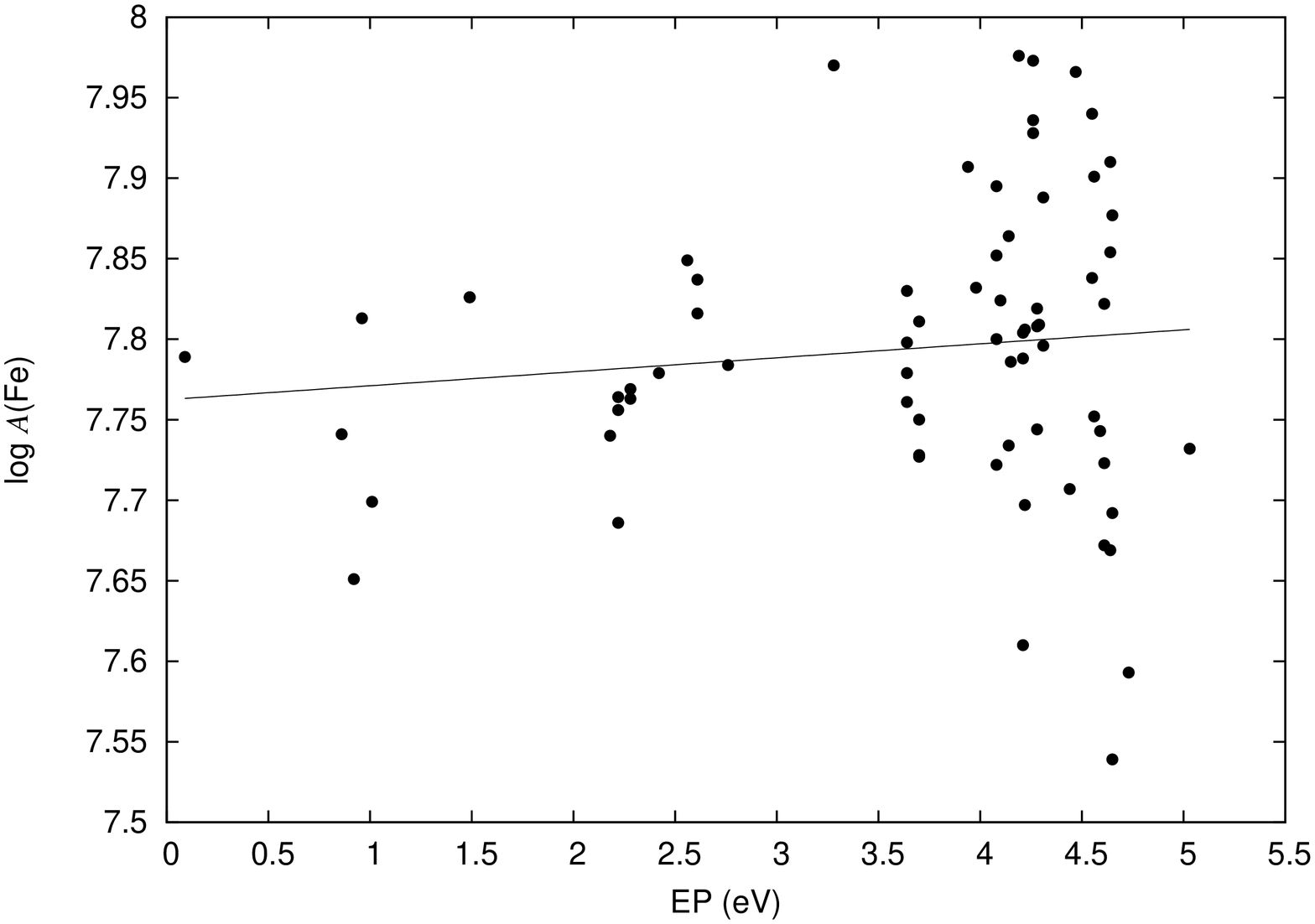}
    \caption{The top panel shows the correct excitation balance at \teff\ = 5197 K for $\alpha$ Cen B using using all of the Fe~{\sc i} lines. The middle panel shows one of the runs with 5 of the Fe~{\sc i} lines removed. The \teff\ derived with this set of lines is 5150 K, however the plot is shown for the correct \teff\ of 5197 K, thus introducing a slope. The bottom panel is for a derived \teff\ = 5249 K, where the 5250 \AA\ line is missing. }
    \label{alphacenB_EP}
\end{figure}

The lowest temperatures (5150 \textendash\ 5185 K) in the plot with 5 Fe~{\sc i} lines removed are caused by the removal of two low EP lines; 6120 and 6625 \AA . The top panel of Figure~\ref{alphacenB_EP} shows the correct excitation balance for a \teff\ of 5197 K using all of the Fe~{\sc i} lines. The middle panel of Figure~\ref{alphacenB_EP} shows the excitation balance for the same \teff, using the set of lines that give a lower \teff\ of 5150 K. The hottest temperatures (5220 \textendash\ 5249 K) are caused by one low EP line (5250 \AA) being removed, as seen in the lower panel of Figure~\ref{alphacenB_EP} for a \teff\ of 5249 K. This shows the importance of including as many low EP lines as possible.

For cool stars, it is extremely difficult to determine the position of the continuum for wavelengths less than $\sim$5200 \AA\ due to the increased strength of the lines, as well as the addition of molecular lines. However, not including the lines less than $\sim$5200 \AA\ only makes the solution worse. This is due to the removal of several low EP Fe~{\sc i} lines, as well as some Fe~{\sc ii} lines. The removal of Fe~{\sc ii} lines, even ones that are difficult to measure, will affect the ionisation balance \logg. In the Sun, removing the lines below 5200 \AA\ results in \logg\ = 4.36 dex.

\section{Discussion}
\label{Discussion}

\subsection{Differential analysis}
When comparing the laboratory line list with the differential line lists, there is good agreement for \teff\ and \logg, however the [Fe/H] of the laboratory line list is systematically higher. In this regard, there is no clear advantage to using the differential lists, especially as it is not possible to tell which metallicity is the ``correct'' value.

When comparing the differential line lists to each other, there is excellent agreement in metallicity between the two lists. However, there is a difference in \teff\ and \logg\ as a function of increasing \teff, most likely due to the different \mictrb. As these lists use different solar \mictrb\ values, the \mictrb\ determined for the stellar sample will also be different and \mictrb\ will affect the \teff\ and \logg.

There is no advantage or disadvantage to using either the laboratory or differential lists when comparing to the bolometric \teff\ or asteroseismic \logg, as the mean difference between the spectral parameters and the external parameters is the same for all lists. However, the mean difference between the differential(1) list and the photometric \teff\ is lower than for the laboratory and differential(0.85) lists. There is excellent agreement between the  \logg\ with the laboratory list and the binary \logg\ of Procyon, but for the other three binary stars, the differential lists are in better agreement.

Using either of the differential lists does not appear to have a negative effect on the parameters for stars with low metallicity or low \logg, which suggests that differential analysis could be used on a wider range of stellar parameters than originally thought, although it would be better to test this with a larger sample.

While the line list using the laboratory \loggf\ values from VALD does result in robust stellar parameters, for future HoSTS papers we will use the differential(1) list as it does seem to produce slightly more accurate results when comparing to the binary \logg\ and the photometric \teff.

\subsection{Fixing \logg\ }
As shown in Figure~\ref{logg_diff_m}, our  spectroscopic \logg\ with the laboratory list does not always agree with the asteroseismic \logg. However, the offset between the values does not agree with that found by either \citet{Bruntt12} or \citet{Mortier14}, showing that while there can be systematic offsets between the $\logg_{\rm spec}$ and $\logg_{\rm ast}$, this depends on the specific method used for the analysis. Our results show that there might be some dependency with \teff, as hotter stars tend to have $\logg_{\rm spec}$ higher than $\logg_{\rm ast}$, where as cooler stars have a lower $\logg_{\rm spec}$. However, the trend is not as obvious as in Figure 6 of \citet{Mortier14}. The mean difference between $\logg_{\rm spec}$ and $\logg_{\rm ast}$ for our data is 0.04 $\pm$ 0.11 dex. If the two outliers, $\delta$ Eri and $\eta$ Boo are removed, this difference reduces to 0.03 $\pm$ 0.08 dex. 

When fixing the \logg\ to the asteroseismic value and redetermining the \teff, we found that the mean difference between the laboratory constrained and unconstrained \teff\ is 3 $\pm$ 13 K, showing that there is excellent agreement between the unconstrained laboratory spectroscopic \teff\ and the spectroscopic \teff\ using the laboratory list obtained when \logg\ is fixed to the asteroseismic value. This is because acquiring the \teff\ via the excitation balance has little or no dependence on the \logg. The excitation balance depends on the EP and the abundance of the Fe~{\sc i} lines, and these lines are not sensitive to changes in surface gravity. Changing the \logg\ will therefore have a minimal effect on the \teff\ derived from this method. 

In contrast, \citet{Mortier14} find a mean difference of 68 K for their sample, however they also take into account the \mictrb\ when determining their temperatures. For this method, after the excitation \teff\ is determined with the \logg\ fixed, the \mictrb\ is adjusted from the Fe~{\sc i} lines only (as the ionisation balance will no longer be met when including the Fe~{\sc ii} lines). Another iteration is run of the excitation balance \teff\ with the new \mictrb. When we computed the laboratory constrained \teff\ using this method, there is a mean difference of 18 $\pm$ 39 K between this and our unconstrained \teff, which is still not as large as that of \citet{Mortier14}.

While fixing \logg\ to the asteroseismic value does not have a significant effect on the spectroscopic parameters determined with this method, it may still be better to use the asteroseismic or transit \logg, if available, when determining the mass and radius of the star as this will result in a more accurate mass and radius for the planet. 

\section{Conclusions}
\label{Conclusion}

In this paper we have used a set of standard stars with known parameters to compare laboratory with differential line lists and also to test if the \logg\ needs to be fixed to an external value. We have shown that the laboratory \loggf\ values from VALD can be used to obtain robust parameters, although there may be a slight advantage to using the differential(1) line list. Despite the fact that we have subgiants and low metallicity stars in our sample, the differential lists still result in robust parameters, showing that it is possible to use the differential line lists across a range of stellar parameters that deviate somewhat from the solar values.

The choice of solar \mictrb\ when creating a differential line list will create a systematic offset in \mictrb\ for the stars analysed. This should be borne in mind when comparing \mictrb\ to that derived via different methods, but it is not a problem when a homogeneous analysis is performed using only one value. The different \mictrb\ will also cause differences in \teff\ and \logg\ as a function of increasing \teff. Using the differential(0.85) list will give higher values of \teff\ and \logg\ for hotter stars compared to the differential(1) list.

The \teff\ is sensitive to the selection of Fe~{\sc i} lines. Randomly removing Fe~{\sc i} lines decreases the chances of returning the correct \teff\ in 1000 temperature iterations. The \teff\ distribution for $\alpha$ Cen A with 5 Fe~{\sc i} lines removed is 5198 $\pm$ 14 K, for 10 Fe~{\sc i} lines removed the \teff\ distribution is 5200 $\pm$ 20 K, and for 30 Fe~{\sc i} lines removed the \teff\ distribution is 5199 $\pm$ 42 K. The \teff\ is particularly sensitive to the number of low EP lines used and removing certain low EP lines can change the \teff\ by $\sim$50 K. Therefore it is important to use as many low EP lines as possible.

We have also shown that fixing the \logg\ to the transit or asteroseismology value offers no improvement to the spectroscopic \teff\ and [Fe/H] for the EW method of spectral analysis. It is therefore sufficient to use the unconstrained values so that the spectral analysis will be self-consistent. However, fixing the \logg\ to an external value does improve the stellar parameters in other methods which rely on comparing the observed spectrum directly to the synthetic spectrum \citep{Torres12}. Also, the transit or asteroseismic \logg\ should be used when determining the stellar mass and radius, as this will improve the precision on the planetary mass and radius.

The inclusion of the \logg\ from the pressure broadened Mg~{\sc i} b and Na~{\sc i} D lines does not improve the result, and the Mg~{\sc i} b \logg\ is usually worse. The Na~{\sc i} D lines can still be used as a check, but this \logg\ is not better than the ionisation balance method.  

While we have chosen to use the differential(1) list and not to fix \logg\ in future HoSTS papers, we conclude that the most important factor is to have a consistent analysis across all stars. Future analyses may also avail of the improved stellar radii, and thus planetary radii, determined from the \textit{Gaia} parallaxes, which will have improved precision after the second data release \citep{Stassun16}.

\section*{Acknowledgements}
We would like to thank Annelies Mortier and Dimitri Veras for helpful discussions. A.P.D. acknowledges support from the Leverhulme grant RPG-2014-082. This work has been partially funded by the Programa UNAM-DGAPA-PAPIIT IA-103215. Based on data obtained from the ESO Science Archive Facility.






\appendix

\section{Line list}
\label{appendix}
\begin{table*}
\centering
\begin{minipage}{170mm}
	\caption{The line list as used in {\sc uclsyn}. The \loggf\ values are are those created differentially to the Sun with \mictrb\ = 1 \kms. The radiative damping constant and the Stark broadening factor are the inverse logs of the values given in VALD, as the inverse log values are required for {\sc uclsyn}. The Van der Waals (VDW) damping constants are listed as they are in VALD for those with ABO values. VDW without ABO values are given as the inverse log. A temperature cutoff is given for some lines if they are too strong or blend above or below a certain \teff. Lines that consistently have an EW > 0.12 \AA\ are deemed to be too strong.}
	\label{tab:example_table}
	\begin{tabular}{lccrcrcl} 
		\hline
Element & Wavelength & EP & \loggf\ & radiative & VDW & stark & \teff\ cutoff \\		
\hline

 Fe~{\sc ii} &  4620.521  &  2.83 & -3.292 & 3.63E+08  &  185.306  & 2.95E-07  &  \\
 Fe~{\sc ii} &  4656.981  &  2.89 & -3.737 & 3.63E+08  &  184.251  & 2.95E-07  &  \\
 Fe~{\sc ii} &  4670.182  &  2.58 & -4.016 & 3.47E+08  &  172.228  & 2.88E-07  &  \\
 Fe~{\sc ii} &  4825.736  &  2.64 & -4.957 & 3.55E+08  &  172.225  & 2.95E-07  &  \\
 Fe~{\sc i}  &  4939.686  &  0.86 &  -3.243 & 1.78E+07  &  244.246  & 7.08E-07  &  \\
 Fe~{\sc ii} &  4993.358  &  2.81 & -3.731 & 3.09E+08  &   172.22  & 2.95E-07  &  \\
 Fe~{\sc i}  &  4994.129  &  0.92 &  -3.186 & 1.74E+07  &  246.245  & 7.08E-07  & Too strong in high [Fe/H] stars < 5800 K \\
 Fe~{\sc ii} &  5000.743  &  2.78 & -4.631 & 3.47E+08  &   173.22  & 2.88E-07  &  Too strong in high [Fe/H] stars < 5300 K\\
 Fe~{\sc i}  &  5016.476  &  4.26 &  -1.647 & 2.40E+08  &  982.279  & 4.07E-05  &  \\
 Fe~{\sc i}  &  5023.198  &  4.28 &  -1.544 & 2.40E+08  & 1013.279  & 1.74E-05  &  \\
 Fe~{\sc i}  &  5058.496  &  3.64 &  -2.773 & 3.39E+07  &  353.313  & 7.94E-07  &  \\
 Fe~{\sc i}  &  5079.740  &  0.99 &  -3.430 & 1.70E+07  &  248.244  & 7.08E-07  &  \\
  Fe~{\sc i}  &  5127.359  &  0.92 &  -3.353 & 1.86E+07  &  243.246  & 7.08E-07  &  \\
 Fe~{\sc i}  &  5127.679  &  0.05 &  -6.072 & 2.88E+03  & 1.51E-08 & 5.25E-07  &  \\
 Fe~{\sc ii} &  5132.669  &  2.81 & -4.120 & 3.47E+08  &  172.219  & 2.95E-07  &  \\
 Fe~{\sc i}  &  5151.911  &  1.01 &  -3.196 & 1.70E+07  &  248.245  & 7.08E-07  & Blended < 5400 K \\
 Fe~{\sc ii} &  5160.839  &  5.57 & -2.000 & 3.02E+08  &  175.234  & 2.95E-07  &  \\
 Fe~{\sc ii} &  5197.577  &  3.23 & -2.300 & 2.88E+08  &  180.247  & 2.95E-07  &  \\
 Fe~{\sc i}  &  5217.389  &  3.21 &  -1.115 & 1.12E+08  &  815.232  & 4.17E-06  &  Too strong < 5800 K\\
  Fe~{\sc i}  &  5247.050  &  0.09 &  -4.973 & 4.27E+03  &  206.253  & 5.25E-07  &  \\
 Fe~{\sc i}  &  5250.209  &  0.12 &  -4.923 & 1.66E+03  &  207.253  & 5.25E-07  & Too strong < 5000 K \\
 Fe~{\sc ii} &  5264.812  &  3.23 & -3.148 & 3.63E+08  &    186.3  & 2.95E-07  &  \\
  Fe~{\sc i}  &  5379.574  &  3.70 &  -1.485 & 7.08E+07  &  363.249  & 7.59E-07  &  \\
 Fe~{\sc ii} &  5414.073  &  3.22 & -3.651 & 3.63E+08  &  185.303  & 2.95E-07  &  \\
 Fe~{\sc i}  &  5421.849  &  4.55 &  -1.938 & 1.62E+08  &  1111.29  & 8.71E-06  &  \\
 Fe~{\sc ii} &  5425.257  &  3.20 &  -3.318 & 2.88E+08  &  178.255  & 2.95E-07  &  \\
 Fe~{\sc ii} &  5427.826  &  6.72 & -1.557 & 3.24E+08  &   173.21  & 2.95E-07  &  \\
  Fe~{\sc i}  &  5441.354  &  4.31 &  -1.594 & 2.04E+08  &  807.278  & 2.00E-05  &  \\
 Fe~{\sc i}  &  5464.278  &  4.14 &  -1.577 & 9.77E+07  & 1.70E-08 & 5.50E-06  &  \\
 Fe~{\sc i}  &  5506.779  &  0.99 &  -2.835 & 1.45E+07  &  241.248  & 6.03E-07  & Too strong < 5800 K \\
 Fe~{\sc i}  &  5522.447  &  4.21 &  -1.409 & 1.05E+08  &  744.215  & 2.69E-06  & Blended < 5800 K \\
 Fe~{\sc i}  &  5538.517  &  4.22 &  -1.536 & 2.82E+08  & 3.31E-08  & 3.39E-05  &  Blended < 4900 K\\
 Fe~{\sc i}  &  5539.284  &  3.64 &  -2.609 & 3.24E+07  &   383.26  & 7.24E-07  &  \\
 Fe~{\sc i}  &  5549.948  &  3.70 &  -2.829 & 3.80E+07  &  373.316  & 8.51E-07  &  \\
 Fe~{\sc i}  &  5560.207  &  4.44 &  -1.095 & 1.91E+08  &  895.278  & 5.75E-05  &  \\
 Fe~{\sc i}  &  5576.090  &  3.43 &  -0.857 & 1.15E+08  &  854.232  & 4.07E-06  & Too strong < 5800 K \\
 Fe~{\sc i}  &  5577.031  &  5.03 &  -1.495 & 7.59E+08  &  4.07E-08 & 5.13E-06  &  \\
 Fe~{\sc i}  &  5607.664  &  4.15 &  -2.228 & 4.07E+08  &  816.278  & 1.32E-05  &  \\
 Fe~{\sc i}  &  5608.974  &  4.21 &  -2.357 & 1.02E+08  &  733.214  & 4.57E-06  &  \\
 Fe~{\sc i}  &  5611.361  &  3.64 &  -2.929 & 1.45E+08  &  376.256  & 1.12E-06  &  \\
 Fe~{\sc i}  &  5618.631  &  4.21 &  -1.311 & 1.55E+08  &  732.214  & 2.69E-06  &  \\
 Fe~{\sc i}  &  5619.224  &  3.70 &  -3.182 & 3.63E+07  &  401.237  & 1.02E-06  &  \\
 Fe~{\sc i}  &  5633.975  &  4.99 &  -0.186 & 7.59E+08  &   635.27  & 8.51E-06  &  \\
 Fe~{\sc i}  &  5635.824  &  4.26 &  -1.602 & 2.29E+08  &  928.279  & 3.39E-05  &  \\
 Fe~{\sc i}  &  5636.696  &  3.64 &  -2.523 & 4.07E+07  &   368.31  & 8.91E-07  &  \\
 Fe~{\sc i}  &  5651.470  &  4.47 &  -1.801 & 1.86E+08  &  898.278  & 3.89E-06  &  \\
 Fe~{\sc i}  &  5652.320  &  4.26 &  -1.800 & 1.02E+08  &   754.21  & 2.69E-06  &  \\
 Fe~{\sc i}  &  5679.025  &  4.65 &  -0.761 & 1.62E+08  & 1106.291  & 8.71E-06  &  \\
 Fe~{\sc i}  &  5680.241  &  4.19 &  -2.349 & 4.90E+07  &  1.70E-08 & 9.33E-07  &  \\
 \hline
 	\end{tabular}
\end{minipage}
\end{table*} 	

 \begin{table*}
\centering
\begin{minipage}{170mm}
	\ContinuedFloat
\caption{Line list (continued)}
	\label{tab:example_table}
	\begin{tabular}{lccrcrcl} 
		\hline
Element & Wavelength & EP & \loggf\ & radiative & VDW & stark & \teff\ cutoff \\		
\hline
 Fe~{\sc i}  &  5724.454  &  4.28 &  -2.549 & 6.46E+07  &  914.278  & 2.04E-05  &  \\
 Fe~{\sc i}  &  5741.846  &  4.26 &  -1.656 & 2.95E+08  &  725.232  & 2.69E-06  &  \\
  Fe~{\sc i}  &  5793.913  &  4.22 &  -1.642 & 2.95E+08  &  714.231  & 5.13E-06  &  \\
 Fe~{\sc i}  &  5806.717  &  4.61 &  -0.928 & 2.09E+08  &  985.281  & 1.95E-05  &  \\
 Fe~{\sc i}  &  5811.917  &  4.14 &  -2.337 & 4.07E+07  & 1.58E-08 & 9.55E-07  &  \\
 Fe~{\sc i}  &  5827.875  &  3.28 &  -3.226 & 1.66E+08  &  748.245  & 3.89E-06  &  \\
 Fe~{\sc i}  &  5849.682  &  3.70 &  -3.011 & 6.03E+07  &  379.305  & 1.32E-06  &  \\
 Fe~{\sc i}  &  5853.150  &  1.49 &  -5.097 & 2.14E+07  & 1.62E-08 & 7.59E-08  &  \\
 Fe~{\sc i}  &  5855.091  &  4.61 &  -1.575 & 2.14E+08  &  962.279  & 2.88E-05  &  \\
 Fe~{\sc i}  &  5856.083  &  4.29 &  -1.572 & 9.77E+07  &  404.264  & 5.50E-06  &  \\
 Fe~{\sc i}  &  5861.107  &  4.28 &  -2.399 & 2.40E+08  &  854.279  & 4.79E-05  &  \\
 Fe~{\sc i}  &  5905.689  &  4.65 &  -0.822 & 2.14E+08  &  994.282  & 1.74E-05  &  \\
 Fe~{\sc i}  &  5929.667  &  4.55 &  -1.237 & 2.14E+08  &  864.275  & 2.63E-05  &  \\
 Fe~{\sc i}  &  5930.173  &  4.65 &  -0.326 & 2.09E+08  &  983.281  & 1.95E-05  & Too strong in high [Fe/H] stars < 5300 K \\
 Fe~{\sc i}  &  5934.653  &  3.93 &  -1.184 & 6.03E+07  &  959.247  & 5.13E-06  &  \\
 Fe~{\sc i}  &  5956.692  &  0.86 &  -4.527 & 1.00E+04  &  227.252  & 6.76E-07  &  \\
 Fe~{\sc ii} &  5991.376  &  3.15 & -3.593 & 3.47E+08  &  172.221  & 2.95E-07  &  \\
 Fe~{\sc i}  &  6012.206  &  2.22 &  -3.845 & 4.79E+06  & 309.27    & 1.67E+02  & Blended > 5800 K \\
 Fe~{\sc i}  &  6024.049  &  4.55 &  0.014 & 2.09E+08  & 823.275   & 8.71E-06  &  Too strong in high [Fe/H] stars < 5800 K\\
 Fe~{\sc i}  &  6027.050  &  4.08 &  -1.073 & 1.02E+08  & 1.66E-08 & 9.33E-07  &  \\
 Fe~{\sc i}  &  6034.033  &  4.31 &  -2.397 & 1.95E+08  & 710.223   & 1.00E-05  &  \\
 Fe~{\sc i}  &  6055.992  &  4.73 &  -0.433 & 2.09E+08  & 1029.286  & 1.12E-05  &  \\
 Fe~{\sc i}  &  6065.482  &  2.61 &  -1.518 & 1.17E+08  & 354.234   & 5.13E-07  & Too strong in high [Fe/H] stars < 5800 K \\
 Fe~{\sc ii} &  6084.111  &  3.20 & -3.815 & 3.47E+08  & 173.223   & 2.95E-07  &  \\
 Fe~{\sc i}  &  6093.666  &  4.61 &  -1.369 & 2.14E+08  & 866.274   & 2.69E-05  &  \\
 Fe~{\sc i}  &  6096.662  &  3.98 &  -1.844 & 5.62E+07  & 963.25    & 5.13E-06  &  \\
 Fe~{\sc i}  &  6098.280  &  4.56 &  -1.786 & 2.75E+08  & 797.269   & 2.00E-05  &  \\
 Fe~{\sc ii} &  6113.322  &  3.22 &  -4.184 & 3.47E+08  & 173.228   & 2.95E-07  &  \\
 Fe~{\sc i}  &  6120.249  &  0.92 &  -5.915 & 1.00E+04  & 1.51E-08 & 6.76E-07  &  \\
 Fe~{\sc ii} &  6149.250  &  3.89 & -2.750 & 3.16E+08  & 186.269   & 2.95E-07  & Blended < 5700 K \\
 Fe~{\sc i}  &  6151.617  &  2.18 &  -3.331 & 1.95E+08  &  277.263  & 6.92E-07  &  \\
 Fe~{\sc i}  &  6157.730  &  4.08 &  -1.135 & 7.76E+07  & 1.62E-08 & 9.33E-07  &  \\
 Fe~{\sc i}  &  6173.340  &  2.22 &  -2.885 & 2.04E+08  &  281.266  & 6.92E-07  &  \\
 Fe~{\sc i}  &  6187.987  &  3.94 &  -1.657 & 5.62E+07  &  903.244  & 4.17E-06  &  \\
 Fe~{\sc i}  &  6200.313  &  2.61 &  -2.330 & 1.20E+08  &  350.235  & 5.13E-07  &  \\
 Fe~{\sc i}  &  6213.429  &  2.22 &  -2.569 & 2.04E+08  &  280.265  & 6.92E-07  &  \\
 Fe~{\sc ii} &  6239.366  &  2.81 & -4.745 & 2.95E+08  &  167.219  & 2.95E-07  & Blended > 5800 K \\
 Fe~{\sc ii} &  6239.953  &  3.89 & -3.481 & 3.16E+08  &  186.271  & 2.95E-07  & Blended > 6000 K \\
 Fe~{\sc i}  &  6240.645  &  2.22 &  -3.309 & 6.46E+06  &  301.272  & 7.41E-07  &  \\
 Fe~{\sc ii} &  6247.557  &  3.89 & -2.383 & 3.16E+08  &  186.272  & 2.95E-07  &  \\
 Fe~{\sc i}  &  6252.554  &  2.40 &  -1.733 & 1.05E+08  &  326.245  & 8.51E-07  & Too strong < 5800 K \\
 Fe~{\sc i}  &  6265.131  &  2.18 &  -2.514 & 2.00E+08  &  274.261  & 6.92E-07  & Too strong 5200 K \\
 Fe~{\sc i}  &  6330.838  &  4.73 &  -1.185 & 2.57E+08  &  915.277  & 3.16E-05  &  \\
 Fe~{\sc i}  &  6335.340  &  2.20 &  -2.254 & 2.00E+08  &  275.261  & 6.92E-07  & Too strong < 5700 K \\
 Fe~{\sc ii} &  6369.462  &  2.89 & -4.188 & 2.95E+08  &  169.204  & 2.95E-07  &  \\
 Fe~{\sc i}  &  6392.538  &  2.28 &  -3.939 & 2.04E+08  &  310.243  & 6.92E-07  &  \\
 Fe~{\sc i}  &  6400.318  &  0.92 &  -4.153 & 2.69E+04  & 1.48E-08 & 6.31E-07  &  \\
 Fe~{\sc ii} &  6432.680  &  2.89 & -3.555 & 2.95E+08  &  169.204  & 2.95E-07  &  \\
 Fe~{\sc ii} &  6446.410  &  6.22 & -2.001 & 4.37E+08  &  181.214  & 2.95E-07  &  \\
 Fe~{\sc ii} &  6456.383  &  3.90 & -2.028 & 3.16E+08  &  185.276  & 2.95E-07  &  \\
 Fe~{\sc ii} &  6482.204  &  6.22 & -1.830 & 3.31E+08  &  181.212  & 2.95E-07  &  \\
 Fe~{\sc i}  &  6498.938  &  0.96 &  -4.629 & 2.29E+04  &  226.253  & 6.17E-07  &  \\
 Fe~{\sc i}  &  6593.871  &  2.43 &  -2.302 & 1.02E+08  &  321.247  & 8.32E-07  & Too strong < 5200 K \\
 Fe~{\sc i}  &  6608.024  &  2.28 &  -3.958 & 2.00E+08  &  306.242  & 6.92E-07  &  \\
 Fe~{\sc i}  &  6625.022  &  1.01 &  -5.340 & 1.15E+04  & 1.48E-08 & 6.17E-07  &  \\
 Fe~{\sc i}  &  6627.540  &  4.55 &  -1.503 & 2.14E+08  &  754.209  & 4.57E-06  &  \\
 Fe~{\sc i}  &  6646.940  &  2.61 &  -3.957 & 9.12E+07  &  339.243  & 8.32E-07  &  \\
 Fe~{\sc i}  &  6699.150  &  4.59 &  -2.125 & 1.23E+08  &  297.273  & 2.34E-06  &  \\
 Fe~{\sc i}  &  6703.568  &  2.76 &  -3.019 & 1.20E+08  &  320.264  & 5.25E-07  &  \\
 \hline
 	\end{tabular}
\end{minipage}
\end{table*} 	

 \begin{table*}
\centering
\begin{minipage}{170mm}
	\ContinuedFloat
\caption{Line list (continued)}
	\label{tab:example_table}
	\begin{tabular}{lccrcrcl} 
		\hline
Element & Wavelength & EP &  \loggf\ & radiative & VDW & stark & \teff\ cutoff \\		
\hline
 Fe~{\sc i}  &  6710.316  &  1.49 &  -4.774 & 1.86E+07  &  252.246  & 7.08E-07  &  \\
 Fe~{\sc i}  &  6725.353  &  4.10 &  -2.209 & 2.51E+08  &  897.241  & 4.17E-06  &  \\
 Fe~{\sc i}  &  6732.070  &  4.58 &  -2.177 & 6.61E+07  &   274.26  & 4.27E-06  &  \\
 Fe~{\sc i}  &  6733.151  &  4.64 &  -1.451 & 2.57E+08  &  781.273  & 4.37E-06  &  \\
 Fe~{\sc i}  &  6745.970  &  4.08 &  -2.711 & 3.09E+07  & 1.51E-08 & 9.77E-07  &  \\
 Fe~{\sc i}  &  6750.150  &  2.42 &  -2.564 & 4.90E+06  &  335.241  & 7.41E-07  &  \\
 Fe~{\sc i}  &  6752.705  &  4.64 &  -1.212  & 2.63E+08  &  778.274  & 3.39E-06  &  \\
 Fe~{\sc i}  &  6806.847  &  2.73 &  -3.112 & 1.17E+08  &  313.268  & 8.13E-07  &  \\
 Fe~{\sc i}  &  6810.257  &  4.61 &  -0.980 & 2.63E+08  &  873.275  & 5.89E-06  &  \\
 Fe~{\sc i}  &  6837.016  &  4.59 &  -1.718 & 7.08E+07  &  273.258  & 7.59E-07  &  \\
 Fe~{\sc i}  &  6839.840  &  2.56 &  -3.333 & 1.12E+08  &  330.248  & 8.71E-07  &  \\
 Fe~{\sc i}  &  6842.679  &  4.64 &  -1.206 & 2.24E+08  &  896.279  & 2.04E-05  &  \\
 Fe~{\sc i}  &  6857.243  &  4.08 &  -2.067 & 1.70E+07  & 1.51E-08 & 9.33E-07  &  \\
 Fe~{\sc i}  &  6858.145  &  4.61 &  -0.953 & 2.69E+08  &  765.211  & 2.69E-06  &  \\
 Fe~{\sc i}  &  6862.492  &  4.56 &  -1.419 & 3.89E+08  &  804.269  & 1.32E-05  &  \\
\hline										

	\end{tabular}
\end{minipage}
\end{table*} 	


\bsp	
\label{lastpage}
\end{document}